\documentclass{aa}  
\usepackage{graphicx}
\usepackage{txfonts}
\usepackage{pdflscape}
\usepackage{soul}
\usepackage{adjustbox}
\usepackage{array}
\usepackage{lscape}

\usepackage{bm}
\usepackage[colorlinks=true,urlcolor=blue,citecolor=blue,linkcolor=blue]{hyperref}

\graphicspath{{./}{figures/}}
\defcitealias{lovdal_2022}{L22}
\defcitealias{dodd_2023}{D23}
\defcitealias{ruiz-lara_2022}{RL22}

\begin{document}

   \title{How well can we unravel the accreted constituents of the Milky Way stellar halo? A test on cosmological hydrodynamical simulations}

   \author{Guillaume F. Thomas\inst{1,2}, Giuseppina Battaglia\inst{1,2}, Robert J. J. Grand\inst{3,1,2} and Amanda Aguiar \'Alvarez\inst{1,2}}

   \institute{Instituto de Astrof\'isica de Canarias, E-38205 La Laguna, Tenerife, Spain \and Universidad de La Laguna, Dpto. Astrofísica, E-38206 La Laguna, Tenerife, Spain \and Astrophysics Research Institute, Liverpool John Moores University, 146 Brownlow Hill, Liverpool, L3 5RF, UK \\
              \email{gthomas@iac.es}
             }

   \date{Submitted on \today}
 \titlerunning{IoM substructures in Auriga}
 \authorrunning{Thomas G. F. et al.}
  \abstract{One of the primary goals of Galactic Archaeology is to reconstruct the Milky Way's accretion history. To achieve this, significant efforts have been dedicated to identifying signatures of past accretion events. In particular, the study of integrals-of-motion (IoM) space has proven to be highly insightful for uncovering these ancient mergers and understanding their impact on the Galaxy's evolution.}{This paper evaluates the effectiveness of a state-of-the-art method for detecting debris from accreted galaxies, by testing it on four Milky Way-like galaxies from the Auriga suite of cosmological magneto-hydrodynamical simulations.}{We employ the innovative method from \citet{lovdal_2022} to identify substructures in the integrals-of-motion space within the local stellar halos of the four simulated galaxies. This approach enables us to evaluate the method's performance by comparing the properties of the identified clusters with the known populations of accreted galaxies in the simulations. Additionally, we investigate whether incorporating chemical abundances and stellar age information can help to link distinct structures originating from the same accretion event.}{This method is very effective in detecting debris from accretion events that occurs less than 6-7 Gyr ago but struggles to detect most of the debris from older accretion. Furthermore, most of the detected structures suffer from significant contamination by in-situ stars. Our results also show that the method may also generate artificial detections.}{Our work show that the Milky Way's accretion history remains uncertain, and question the reality of some detected structures in the Solar vicinity.}


   \maketitle
%

\section{Introduction}

In the current $\Lambda$  Cold Dark Matter ($\Lambda$CDM) cosmological model, the accretion of smaller satellite galaxies and interactions between galaxies play a fundamental role in the formation and evolution of galaxies. Indeed, large galaxies, such as the Milky Way (MW), grow hierarchically by accreting gas from cosmic filaments, which drives their secular evolution and leads to the formation of in-situ stars, but also by cannibalizing smaller satellite galaxies during successive accretion events \citep{eggen_1962,searle_1978,white_1991,springel_2005,carollo_2007,carollo_2010,purcell_2011,qu_2017}. Therefore, it is fundamental to understand the relative importance of these two formation channels to shape the properties of galaxies as we see them today.  Furthermore, determining the epoch of accretion of the accreted galaxies, and studying their internal properties—such as mass, luminosity, chemical evolution, star formation history, and associations with globular clusters or other galaxies—are essential for constraining cosmological models. However, the majority of the accretion events occurred in the early Universe \citep[e.g.][]{blumenthal_1984,grand_2017,monachesi_2019,horta_2023}. As such, observing this process while it is ongoing at high redshift will likely remain unfeasible in the foreseeable future due to their inherently faint luminosity, even with cutting-edge instruments like the James Webb Space Telescope (JWST). Although, it has recently been shown that the JWST can detect galaxy pairs prior to their merger for systems with stellar masses as low as 10$^8$ M$_\odot$ \citep{duan_2024}, its use in studying these mergers remains limited. Therefore, the Galactic Archaeology approach of uncovering past accretion and merger events by identifying the stars shed among those of the host galaxy is fundamental to complement work done with large galaxy samples, including those at high redshifts \citep[e.g.][]{conselice_2014,costantin_2023}, in order to develop a comprehensive understanding of the processes driving the formation and evolution of galaxies across different environments and epochs.

Over the lifetime of a large galaxy, the spatial coherence of an accreted galaxy is rapidly lost due to phase-mixing, especially for the most massive accreted galaxies. These massive galaxies, due to dynamical friction, quickly sink into the centre of the host, where the dynamical timescale is on the order of a few hundred million years \citep{amorisco_2017,vasiliev_2022}. Even less massive galaxies situated in the outer stellar halo—where ongoing galaxy disruption can be observed in the form of stellar streams \citep[see][and references therein]{amorisco_2017,mateu_2023,ibata_2024}—typically lose their spatial coherence within a few billion years \citep{johnston_2008,gomez_2010a}. 

Despite this loss of spatial coherence, it was postulated that the remnants of accreted galaxies remain clustered in integrals-of-motion (IoM) space or in action space, which would allow to still identify the debris left by past accreted galaxies 
\citep{helmi_1999,helmi_2000,knebe_2005,brown_2005,font_2006,mcmillan_2008,morrison_2009,gomez_2010}. Additionally, the chemical composition of individual stars serves as an additional tool for identifying the remnants of these accreted galaxies, as stars from the same galaxy share similar chemical abundance patterns \citep{freeman_2002,venn_2004,lee_2015,fernandes_2023}, which are somewhat distinct from those of stars formed in situ \citep[e.g.][]{hawkins_2015,haywood_2018,das_2020,horta_2021,belokurov_2022}. 

In this context, the Milky Way offers a unique opportunity to study the accretion history of an individual galaxy. Not only is it a typical galaxy within the Local Universe \citep{kormendy_2010,vandokkum_2013,papovich_2015,bland-hawthorn_2016}, but it is also the only massive galaxy for which we can obtain full 6D phase-space information, detailed chemical abundances, and relative ages for large samples of individual stars. Although similar attempts have been made on other nearby galaxies within the Local Volume \citep[e.g.][]{gilbert_2014,dsouza_2018,mcconnachie_2018,mackey_2019,zhu_2020,davison_2021}, the Milky Way remains unparalleled in the level of detail that can be achieved.

Thanks to the ESA flagship mission Gaia \citep{gaiacollaboration_2018}, which provides accurate astrometry, parallaxes, proper motions, and even radial velocities and stellar parameters and chemical abundance \citep{recio-blanco_2023}, complemented with ground-based spectroscopic surveys that provide radial velocities and chemical abundances for stars too faint for Gaia RVS —such as SDSS/SEGUE \citep{yanny_2009}, LAMOST \citep{zhao_2012a,yan_2022}, RAVE \citep{steinmetz_2006,steinmetz_2020,steinmetz_2020a}, GALAH \citep{buder_2021}, APOGEE \citep{abdurrouf_2022}, Gaia-ESO \citep{gilmore_2022,randich_2022}, and soon-to-be complemented by new surveys like DESI-MWS \citep{cooper_2023}, WEAVE \citep{dalton_2012,jin_2024}, 4-MOST \citep{dejong_2019}, and SDSS-V \citep{kollmeier_2017}— it is now possible to detect the dynamical debris left by past accreted galaxies, even if they do not form spatially coherent structures anymore (see \citealt{helmi_2020} and \citealt{deason_2024} for recent reviews). 

With these data, numerous stellar debris associated with accreted galaxies have been discovered in the local stellar halo complementing already known debris, such as the Sagittarius dwarf galaxy \citep[M$_\star = 10^8$ M$_\odot$ accreted $\sim$4-6~Gyr ago;][]{ibata_1994,majewski_2003,niederste-ostholt_2010}. These includes Gaia-Enceladus-Sausage \citep[GES; M$_\star = 10^8-10^{10}$ M$_\odot$ accreted $\sim$10~Gyr ago;][]{helmi_2018,vincenzo_2019,mackereth_2019,gallart_2019,feuillet_2020,naidu_2022,lane_2023}, Heracles-Kraken-IGS \citep[M$_\star \sim 2\times 10^8$ M$_\odot$ accreted $\sim$11~Gyr ago;][]{kruijssen_2019,kruijssen_2020,massari_2019,horta_2021}, Sequoia \citep[M$_\star \sim 5\times10^7$ M$_\odot$ accreted $\sim$9~Gyr ago;][]{myeong_2018,myeong_2019,naidu_2020,matsuno_2022}, the Helmi stream \citep[M$_\star \sim 10^8$ M$_\odot$, accreted $5-9$~Gyr ago;]{helmi_1999,kepley_2007,koppelman_2019}, and Thamnos \citep[M$_\star \sim 5\times 10^6$ M$_\odot$, accreted $>$10~Gyr ago;][]{koppelman_2019b,ruiz-lara_2022,dodd_2024}. This list is far from exhaustive, and many recent studies have also identified smaller stellar substructures \citep[e.g., LMS-1/Wukong, Pontus, Typhon/ED-4, ED-2-6, Shakti, Shiva, Rg5, Arjuna, L'Itoi, Nyx, L-RL64/Antaeus, etc...;][]{malhan_2021,malhan_2022,naidu_2020,necib_2020,yuan_2020,myeong_2022,oria_2022,tenachi_2022,lovdal_2022,ruiz-lara_2022,dodd_2023}, though the accreted origin of some remains under debate \citep[e.g., Nyx, Aleph,][]{necib_2020,naidu_2020,zucker_2021,horta_2023}. 

In most cases, the detection of these structures is carried out either through manual selection \citep[e.g.,][]{naidu_2020,oria_2022,tenachi_2022} or by using clustering algorithms in dynamical or chemical space \citep[e.g.,][]{koppelman_2019,myeong_2022}. However, the selection criteria used in manual methods, or the choice of parameters in clustering algorithms, can significantly influence which stars are assigned to a particular structure and may even alter the perceived properties of the structures themselves \citep{rodriguez_2019,carrillo_2024}. Machine learning techniques have also been employed to find clusters in IoM or action space \citep{yuan_2018, myeong_2018c, borsato_2020,shih_2022}, and have successfully identified various stellar streams. They were also used to separate accreted stars, or globular clusters \citep{trujillo-gomez_2023}, from those formed in-situ \citep{veljanoski_2019,ostdiek_2020,tronrud_2022,sante_2024}. However, all these different methods have their one flow and limitation, and in particular none is currently capable of measuring robustly the significance of each of these detections.

More recently, \citet{lovdal_2022} (hereafter \citetalias{lovdal_2022}) introduced a  data-driven algorithm designed to detect and evaluate the significance of substructures in IoM space. This method is based on a single-linkage clustering approach and quantifies the number of stars grouped together by comparing them against a set of artificial background halos. What sets this technique apart is its ability to not only identify substructures but also to assign a robust statistical significance to each detection, while simultaneously providing insights into potential associations between distinct substructures by incorporating  additional information such as the metallicity or the colour distribution. The effectiveness of this method has been demonstrated by its successful identification of new substructures in the IoM space within the local stellar halo of the Milky Way \citep[][hereafter \citetalias{ruiz-lara_2022} and \citetalias{dodd_2023}, respectively]{ruiz-lara_2022,dodd_2023}.

Another aspect to be considered is that the {\it image d'\'Epinal} of the preservation of the phase-space properties of the debris left by accreted galaxies and their significant chemical difference with in-situ stars needs to be somewhat relativised. Cosmologically motivated simulations have demonstrated that a single accreted galaxy can give rise to multiple distinct structures in IoM or action space, which may further exhibit varying chemical distributions due to the underlying metallicity gradient in the original galaxy \citep{jean-baptiste_2017, grand_2019, amarante_2022, khoperskov_2023, khoperskov_2023a, mori_2024}. Conversely, the same simulations suggest that a given structure might result from the accumulation of stellar debris from multiple accretions or could be contaminated by in-situ stars \citep{naidu_2020, orkney_2022, khoperskov_2023a}. Similar complexities have also been observed in groups of globular clusters \citep{pagnini_2023}.

In this paper, we assess the effectiveness of the \citetalias{lovdal_2022} method in identifying structures and determining their significance using solar-neighbourhood-like stellar mocks derived from four Milky Way-like galaxies from the Auriga suite of cosmological magneto-hydrodynamic simulations \citep{grand_2017,grand_2024}. In Sect.~\ref{sec:data}, we present the properties of the primary progenitors of these four simulated galaxies. Sect.~\ref{sec:method} describes the creation of the stellar halo mocks and introduces the \citetalias{lovdal_2022} algorithm, including the modifications made to adapt it for use with the different mocks. The results of applying the algorithm to the mocks are analysed in Sect.~\ref{sec:results}. The properties of significant clusters are detailed in Sect.~\ref{sec:recovrate}, and their chemical and age characteristics are thoroughly examined in Sect.~\ref{sec:chemage}. In Sect.~\ref{sec:GroupClumps}, we investigate the purity and completeness of groups of dynamically close clusters, while Sect.~\ref{sec:acconly} evaluates the algorithm's effectiveness in scenarios lacking in-situ stars. Sect.~\ref{sec:backgen} explores the impact of artificial background halos on the significance of the detected clusters. Finally, in Sect.~\ref{sec:conclusion}, we summarize our findings, discussing both the challenges inherent in the methods used for identifying and quantifying substructures, and the more intrinsic issues stemming from the initial expectations regarding the distribution in phase space and the chemical abundance patterns trends of accreted galaxies. We then use these insights to propose hypothetical properties of the accreted structures identified in the Milky Way to date.

\section{Data} \label{sec:data}

\begin{figure*}
\centering
  \includegraphics[angle=0,clip,width=18cm]{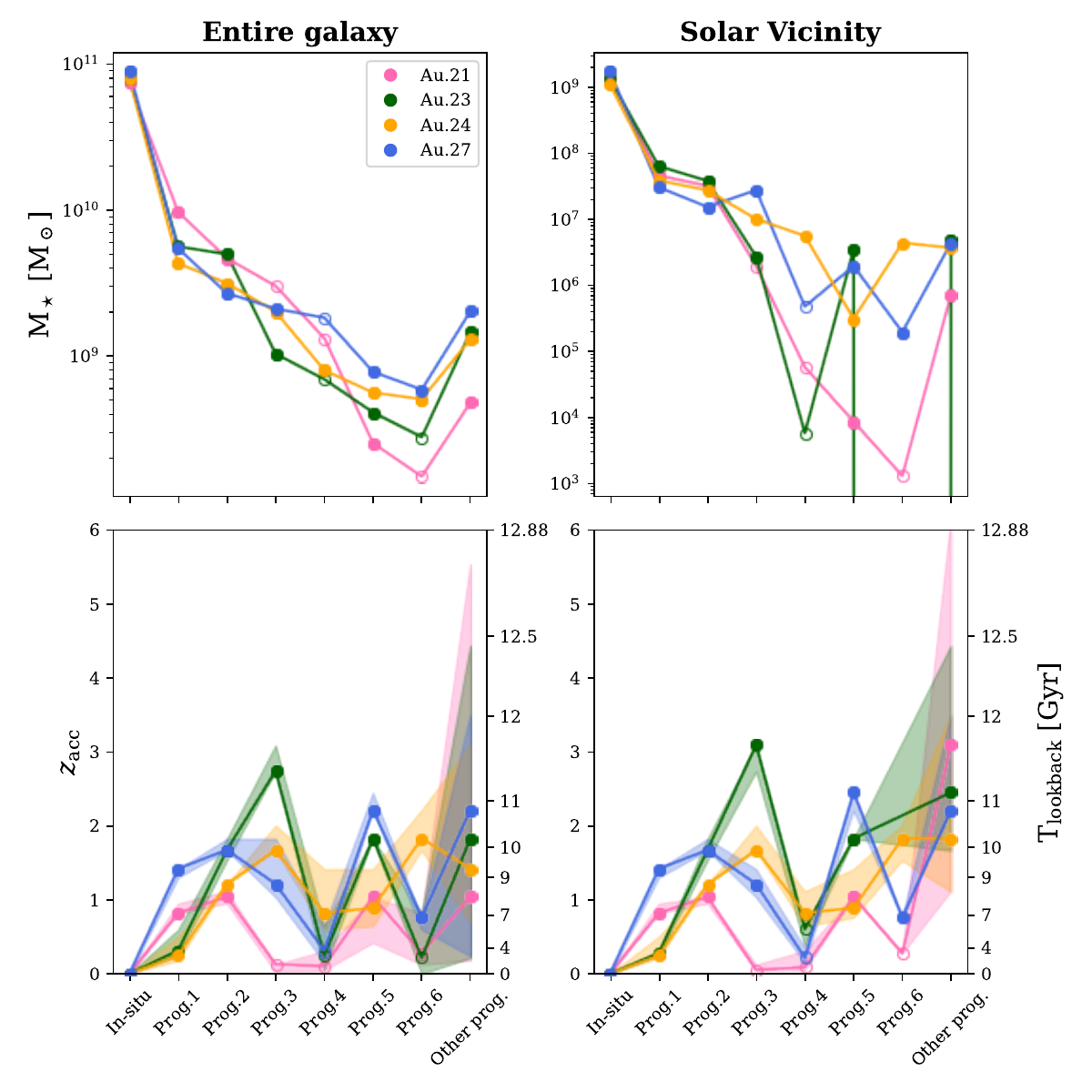}
   \caption{Current ($z=0$) stellar mass (upper row) and redshift of accretion (lower row) of the six most massive accreted progenitors of the four Auriga MW-like galaxies studied here. The stellar mass of the in-situ component at $z=0$ is also indicated for comparison, as it is the redshift of accretion of the smaller progenitors. The left column displays these properties as calculated within the entire galaxy (i.e. $<R_{200}$), while the right column for the Solar vicinity region (see Section~\ref{sec:haloSun}). In the bottom panels, the circles (shaded regions) indicate the (range of) redshift when 50\% (5-90\%) of the stellar particles from a progenitor were accreted by the main galaxy. The accreted galaxies with a surviving progenitor with more than 1\% of the initial mass of the progenitor at $z=0$ are indicated by an open circle.} 
\label{fig:massZprog}
\end{figure*}

\begin{figure*}
\centering
  \includegraphics[angle=0,clip,viewport=0 100 1000 465,width=18cm]{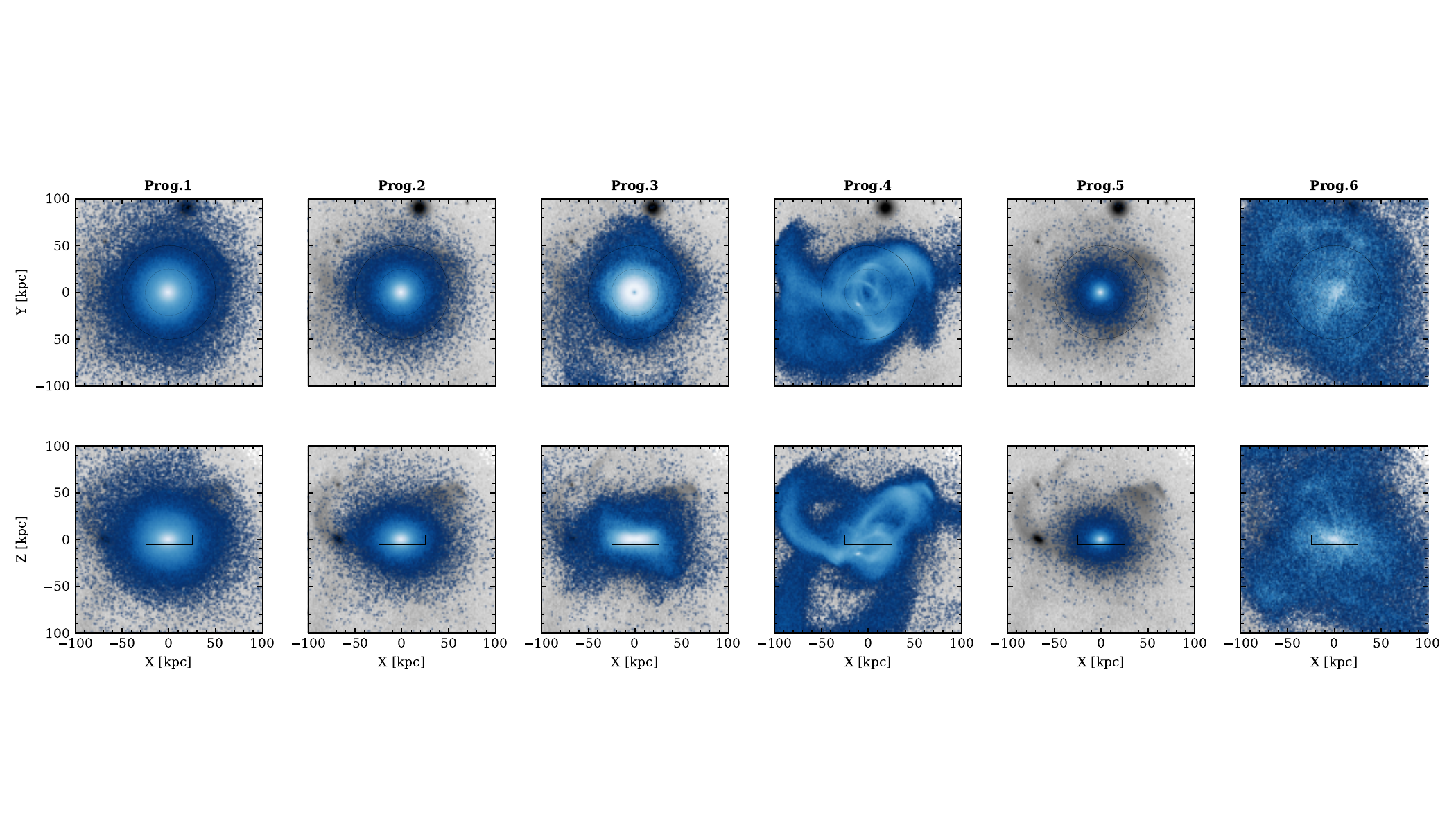}
   \caption{Spatial distribution at $z=0$ of the particles (accreted or still bound to the progenitor) for the six most massive progenitors (blue) compared to all particles (grey) in Au.~27.  Note that the physical distances presented here have been rescaled, such as the scale-length of the simulated disc is similar to the scale-length of the MW (see Section~\ref{sec:overview}).} 
\label{fig:St_M_D}
\end{figure*}

\begin{table*}[!h]
\centering
\caption{Properties of the six most massive accreted galaxies and of the in-situ component at $z$=0 for the four MW-like galaxies from the Auriga simulations studied here. All the less massive progenitors are regrouped together under the label``Other prog.''. Col.~2 indicates the ration between the total mass of the progenitor and the host galaxy at infall time, and the value in parentheses corresponds to the ratio between the stellar mass of the progenitor and the host. Col.~3 shows the stellar mass of each progenitor/in-situ component in the entire volume of the galaxy (<$R_{200}$). Col.~4 shows their stellar mass contribution in a sphere of 2.5~kpc radius around the Sun (solar vicinity).  In these columns, the percentage of the stellar contribution of each progenitor relative to the total stellar mass in the respective volume is provided in parentheses. Col.~5 indicates the redshifts when 5, 50, and 90\% of the particles of a given progenitor escaped its gravitational attraction and became bound to the main galaxy. Col.~6 indicated if after a visual inspection, the spatial distribution of the debris of an accreted galaxy present a stream-like (St) or a disc-like (D) structure, or if they do not display any particular shape and are then fully spatially-mixed (M).}
\label{tab:param}
\begin{tabular}{llcccc}
\multicolumn{6}{c}{\bf{Halo 21}}\\
\hline
Progenitor & Merger ratio & M$_\star (<R_{200})$ [$10^{9}$ M$_\odot$]& M$_\star (R_{helio}<2.5$~kpc$)$  [$10^{7}$ M$_\odot$] & $z_{acc,5-50-90}$ & State  \\
\hline
\bf{Total} & & \bf{92.30} & \bf{144.609} & \\
In-situ & & 72.88 ($79.0 \%$) & 136.789 ($94.6 \%$)& & \\
Prog.~1 & 6:1 (2:1) & 9.58 ($10.4 \%$) & 4.454 ($3.1 \%$) & [1.0, 0.8, 0.8] & M \\
Prog.~2 & 8:1 (2.5:1) & 4.64 ($5.0 \%$) & 3.100 ($2.1 \%$) & [1.1, 1.0, 1.0] & M \\
Prog.~3$^*$ & 20:1 (25:1) & 3.00 ($3.2 \%$) & 0.191 ($0.1 \%$) &[0.1, 0.1, 0.0] & St\\
Prog.~4$^*$ & 12:1 (11:1) & 1.32 ($1.4 \%$) & 0.006 ($0.0 \%$) & [0.3, 0.1, 0.0] & St \\
Prog.~5 & 36:1 (40:1) & 0.25 ($0.3 \%$) & 0.001 ($0.0 \%$) & [1.0, 1.0, 0.4]  & St\\
Prog.~6$^*$ & 43:1 (57:1) & 0.15 ($0.2 \%$) & 0.000 ($0.0 \%$) & [0.8, 0.3, 0.1] & St \\
Other prog. & & 0.048 ($0.5 \%$) & 1.0 ($0.1 \%$) & [5.5, 1.0, 0.2] & \\
\hline\\

\multicolumn{6}{c}{\bf{Halo 23}}\\
\hline
Progenitor &  Merger ratio & M$_\star (<R_{200})$ [$10^{9}$ M$_\odot$] & M$_\star (R_{helio}<2.5$~kpc$)$ [$10^{7}$ M$_\odot$] & $z_{acc,5-50-90}$ & State  \\
\hline
\bf{Total} & & \bf{93.33} & \bf{138.237} & \\
In-situ & & 78.95 ($84.6 \%$) & 127.304 ($92.1 \%$) & & \\
Prog.~1 & 16:1 (15:1) & 5.57 ($6.0 \%$) & 6.154 ($4.5 \%$) & [0.6, 0.3, 0.2] & D \\
Prog.~2 & 22:1 (5:1) & 4.94 ($5.3 \%$) & 3.687 ($2.7 \%$) & [1.8, 1.7, 1.5] & M \\
Prog.~3 & 7:1 (3:1) & 1.03 ($1.1 \%$) & 0.265 ($0.2 \%$) & [3.1, 2.7, 2.7] & M \\
Prog.~4$^*$ & 53:1 (84:1) & 0.69 ($0.7 \%$) & 0.001 ($0.0 \%$) & [0.6, 0.2, 0.0] & St \\
Prog.~5 & 114:1 (41:1) & 0.41 ($0.4 \%$) & 0.338 ($0.2 \%$) & [1.8, 1.8, 1.8] & M \\
Prog.~6$^*$ & 47:1 (121:1) & 0.28 ($0.3 \%$) & 0.000 ($0.0 \%$) & [0.8, 0.2, 0.0] & St \\
Other prog. &  & 1.46 ($1.6 \%$) & 0.489 ($0.4 \%$) & [4.4, 1.8, 0.2] & \\
\hline \\

\multicolumn{6}{c}{\bf{Halo 24}}\\
\hline
Progenitor & Merger ratio & M$_\star (<R_{200})$ [$10^{9}$ M$_\odot$] & M$_\star (R_{helio}<2.5$~kpc$)$ [$10^{7}$ M$_\odot$] & $z_{acc,5-50-90}$ & State  \\
\hline
\bf{Total} & & \bf{89.89} & \bf{114.400} & \\
In-situ & & 77.32 ($86.0 \%$) & 105.676 ($92.4 \%$) & & \\
Prog.~1 & 18:1 (17:1) & 4.27 ($4.8 \%$) & 3.729 ($3.3 \%$) & [0.5, 0.2, 0.2] & D \\
Prog.~2 & 10:1 (8:1) & 3.13 ($3.5 \%$) & 2.655 ($2.3 \%$) & [1.2, 1.2, 1.2] & M \\
Prog.~3 & 51:1 (4:1) & 2.00 ($2.2 \%$) & 0.977 ($0.9 \%$) & [2.0, 1.7, 1.7] & M \\
Prog.~4 & 29:1 (23:1) & 0.80 ($0.9 \%$) & 0.545 ($0.5 \%$) & [1.4, 0.8, 0.6] & D \\
Prog.~5 & 64:1 (40:1) & 0.56 ($0.6 \%$) & 0.030 ($0.0 \%$) & [1.4, 0.9, 0.6] & St \\
Prog.~6 & 22:1 (14:1) & 0.51 ($0.6 \%$) & 0.427 ($0.4 \%$) & [2.2, 1.8, 1.7] & M \\
Other prog. & & 1.30 ($1.4 \%$) & 0.360 ($0.3 \%$) & [3.1, 1.4, 0.7] & \\
\hline \\

\multicolumn{6}{c}{\bf{Halo 27}}\\
\hline
Progenitor & Merger ratio & M$_\star (<R_{200})$ [$10^{9}$ M$_\odot$] & M$_\star (R_{helio}<2.5$~kpc$)$ [$10^{7}$ M$_\odot$] & $z_{acc,5-50-90}$ & State  \\
\hline
\bf{Total} & & \bf{102.49} & \bf{172.826} & \\
In-situ & & 87.12 ($85.0 \%$) & 165.019 ($95.5 \%$) & & \\
Prog.~1 & 9:1 (4:1) & 5.39 ($5.3 \%$) & 2.949 ($1.7 \%$) & [1.4, 1.4, 1.3] & M \\
Prog.~2 & 9:1 (4:1)& 2.68 ($2.6 \%$) & 1.493 ($0.9 \%$) & [1.8, 1.7, 1.7] & M \\
Prog.~3 & 55:1 (8:1) & 2.10 ($2.0 \%$) & 2.680 ($1.6 \%$) & [1.8, 1.2, 1.0] & D \\
Prog.~4$^*$ & 28:1 (56:1)& 1.82 ($1.8 \%$) & 0.047 ($0.0 \%$) & [0.6, 0.3, 0.2] & St \\
Prog.~5 & 44:1 (4:1)& 0.78 ($0.8 \%$) & 0.191 ($0.1 \%$) & [2.4, 2.2, 2.2] & M \\
Prog.~6 & 53:1 (86:1) & 0.59 ($0.6 \%$) & 0.019 ($0.0 \%$) & [0.8, 0.8, 0.6] & M \\
Other prog. & & 2.02 ($2.0 \%$) & 0.429 ($0.2 \%$) & [3.5, 2.2, 0.2] & \\
\hline

\end{tabular}
\tablefoot{Note that we indicate here only the stellar mass of the star particles of each progenitor flagged as {\it accreted} ({\sc AccretedFlag$==0$}). This does not take into account the mass of the stellar particles that are still bound to the satellite galaxy. The asterix after the progenitor id indicate that more than 1\% of the stellar mass of the accreted galaxy are still bound the the surviving progenitor. Moreover, the mass of each progenitor in the Solar vicinity (Col. 3) correspond to the average mass in the Solar vicinity of the four azimuths used to make the mocks.}
\end{table*}

\begin{figure*}
\centering
  \includegraphics[angle=0,clip,width=17cm]{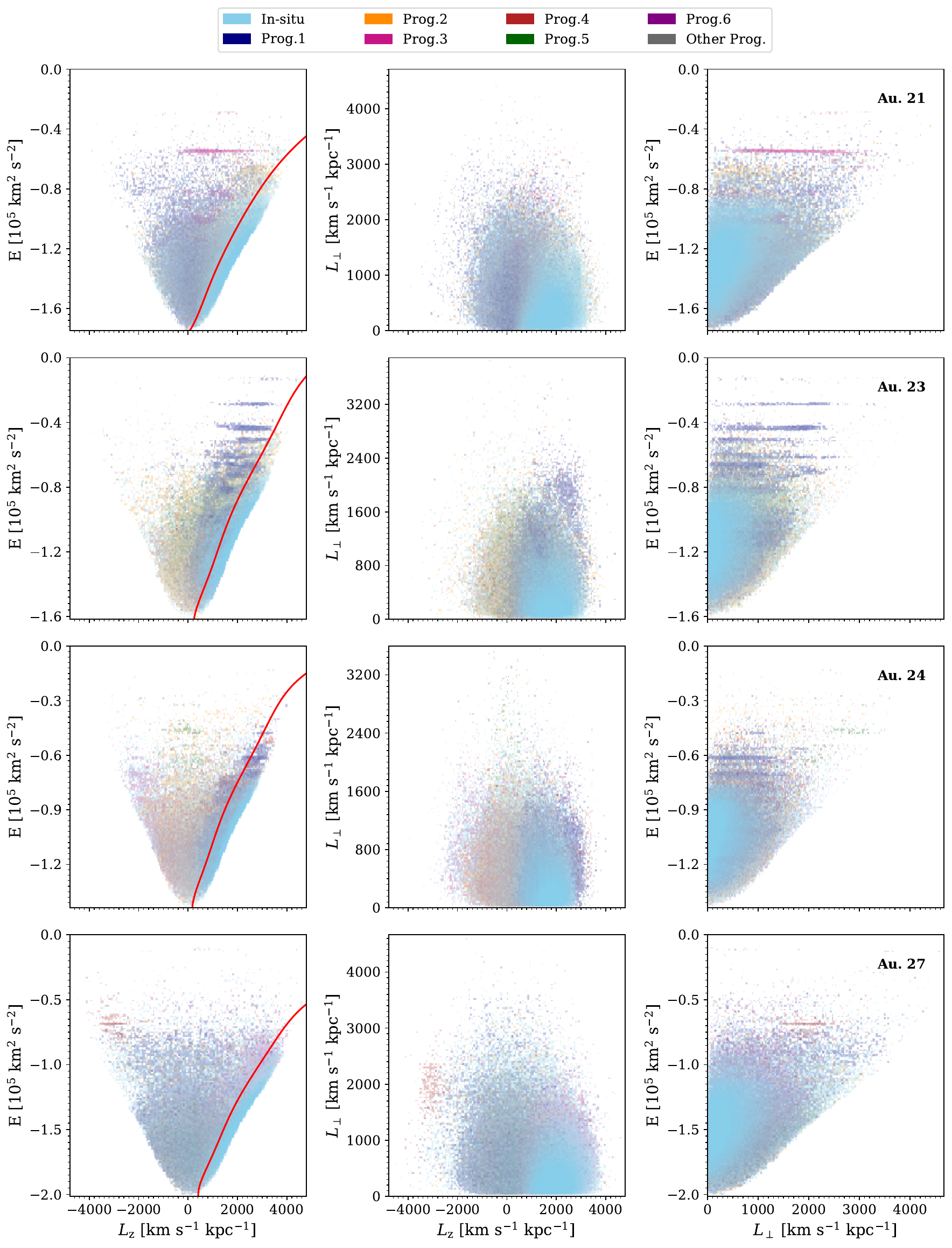}
   \caption{Dominance diagrams of integrals of motion space of the Solar vicinity for the four simulated galaxies analysed. Each progenitor is identified by a specific colour, as indicated in the legend. In these diagrams, the local uniqueness of a colour indicates that the region is mainly populated by particles originating from a single progenitor. On the contrary, regions where particles have different origins are identified by areas with blended colours, proportional to the local contribution of each progenitor. The density variation is logarithmically proportional to the colour opacity. The red line on the left panels {\it indicates} the approximate limit of the kinematically selected halo (see Section~\ref{sec:haloSun}).} 
\label{fig:dominance}
\end{figure*}

\subsection{Sample of Milky Way-like galaxies} \label{sec:overview}
We perform our analysis on a sample of high resolution gravo-magnetohydrodynamic cosmological zoom-in simulations of Milky Way-mass halos ($1 < M_{200}/[10^{12}\, {\rm M_{\odot}}] < 2$ at redshift 0\footnote{Here, $M_{200}$ is defined as the mass contained inside the radius, $R_{200}$, at which the mean enclosed mass volume density equals 200 times the critical density for closure.}) taken from the Auriga project \citep{grand_2017}. These simulations adopt the following cosmological parameters: $\Omega _m = 0.307$, $\Omega _b = 0.048$, $\Omega _{\Lambda} = 0.693$, $\sigma_8 = 0.8288$, and a Hubble constant of $H_0 = 100\, h\, \rm km \, s^{-1} \, Mpc^{-1}$, where $h = 0.6777$, taken from \citet{planckcollaboration_2014}. The initial conditions are generated for a starting redshift of 127, and follow the evolution of gas, dark matter, stars, and black holes down to redshift 0 according to a comprehensive galaxy formation model. The model includes: primordial and metal-line radiative cooling and heating from a spatially uniform, redshift-dependent Ultra Violet background radiation field with self-shielding corrections \citep{faucher-giguere_2009}; an effective 2-phase sub-grid model for the multi-phase interstellar medium (ISM) \citep{springel_2003}; stochastic star formation in dense ISM gas above a threshold density of $n = 0.13$ $\rm cm^{-3}$ assuming a \citet{chabrier_2003} initial mass function; stellar evolution including mass-loss and chemical enrichment of surrounding gas from asymptotic giant branch (AGB) stars, supernovae Ia and II; an energetic stellar feedback scheme that models galactic-scale gaseous outflows; seeding and growth of supermassive black holes via Bondi accretion; thermal feedback from active galactic nuclei (AGNs) in quasar and radio modes. A full description of the simulations can be found in \citet{grand_2017}.

In this study, we focus on the high resolution ``level 3'' simulations: each star particle/gas cell is $\sim 6\times 10^3$ $\rm M_{\odot}$. These are the highest resolution Auriga simulations available, and therefore maximise the amount of stellar substructure predicted from the cosmological assembly of Milky Way-mass haloes. Specifically, we select 4 galaxies (Au.~21, Au.~23, Au.~24, Au.~27), as they have either experienced the radial accretion (with an anisotropy $\beta>0.7$) of a satellite galaxy that contributed at least 40\% of the stellar halo mass \citep[see Figure~3 of][]{fattahi_2019}, akin to the properties of the progenitor of the Gaia-Enceladus-Sausage \citep{helmi_2018,belokurov_2018} accreted 8-11 Gyr ago by the MW \citep{dimatteo_2019,gallart_2019}, or they have a massive satellite galaxy on a first infall orbit, similar to the leading hypothesis for the Large Magellanic clouds (\citealt{smith-orlik_2023}; but see \citealt{vasiliev_2024} for a 2-passage scenario). The simulations for these objects are publicly available\footnote{\url{https://wwwmpa.mpa-garching.mpg.de/auriga/gaiamock.html}} \citep{grand_2018,grand_2024}.

As noticed in previous works, the disc of the simulated Auriga galaxies are more extended than the disc of the MW, and present a wide diversity of radial scale-length, ranging from 2.16 to 11.64 kpc \citep{grand_2017}. In comparison, the scale-length of the MW disc is $R_\textrm{d,MW}=2.6 \pm{0.5}$ kpc  \citep[average value from 15 literature studies compiled by][]{bland-hawthorn_2016}\footnote{it has to be noted that the scale-length measured depend on the age of the observed population, with younger population being more extended than the older one \citep{bovy_2012a}}. The 4 simulated galaxies used in this paper are no exception, since their scale-length varies from 3.2 kpc (Au.~27) to 6.1 kpc \citep[Au.~24,][]{grand_2018}. As a consequence, at a physical radius of 8 kpc the density of stars from the disc, which are mostly formed in-situ, is more important than in the MW. Therefore, to ensure that the simulations are comparable to the MW in the Solar vicinity, we scale each system—unless explicitly stated otherwise—such that the Solar volume is located at the same position w.r.t the disc scale length (approximately 3 times the scale radius), matching the value measured in the MW.

These re-scaled galaxies are then used to produce mock samples of stellar particles having similar properties as the Gaia sample used by \citetalias{lovdal_2022} and \citetalias{ruiz-lara_2022} to search for the signature of accreted galaxies in the MW stellar halo around the Solar neighbourhood. We build these Solar neighbourhood-like mocks by selecting stellar particles located within four spheres, of radius 2.5 kpc, centred at the Sun Galactocentric distance of 8.129 kpc \citep{gravitycollaboration_2018} and placed at four different azimuths spaced by 90 degrees. The choice of having multiple spheres is driven by the need of having halo-like mocks with the same ballpark number of stellar particles as the number of stars with halo-like kinematics in \citet[][$\sim 69,000$, in the former versus $87,000-119,000$ for the Auriga Solar neighbourhood mocks, see Sect.~\ref{sec:haloSun}]{dodd_2023}.

\begin{figure*}[htb!]
\centering
  \includegraphics[angle=0,clip,width=17cm]{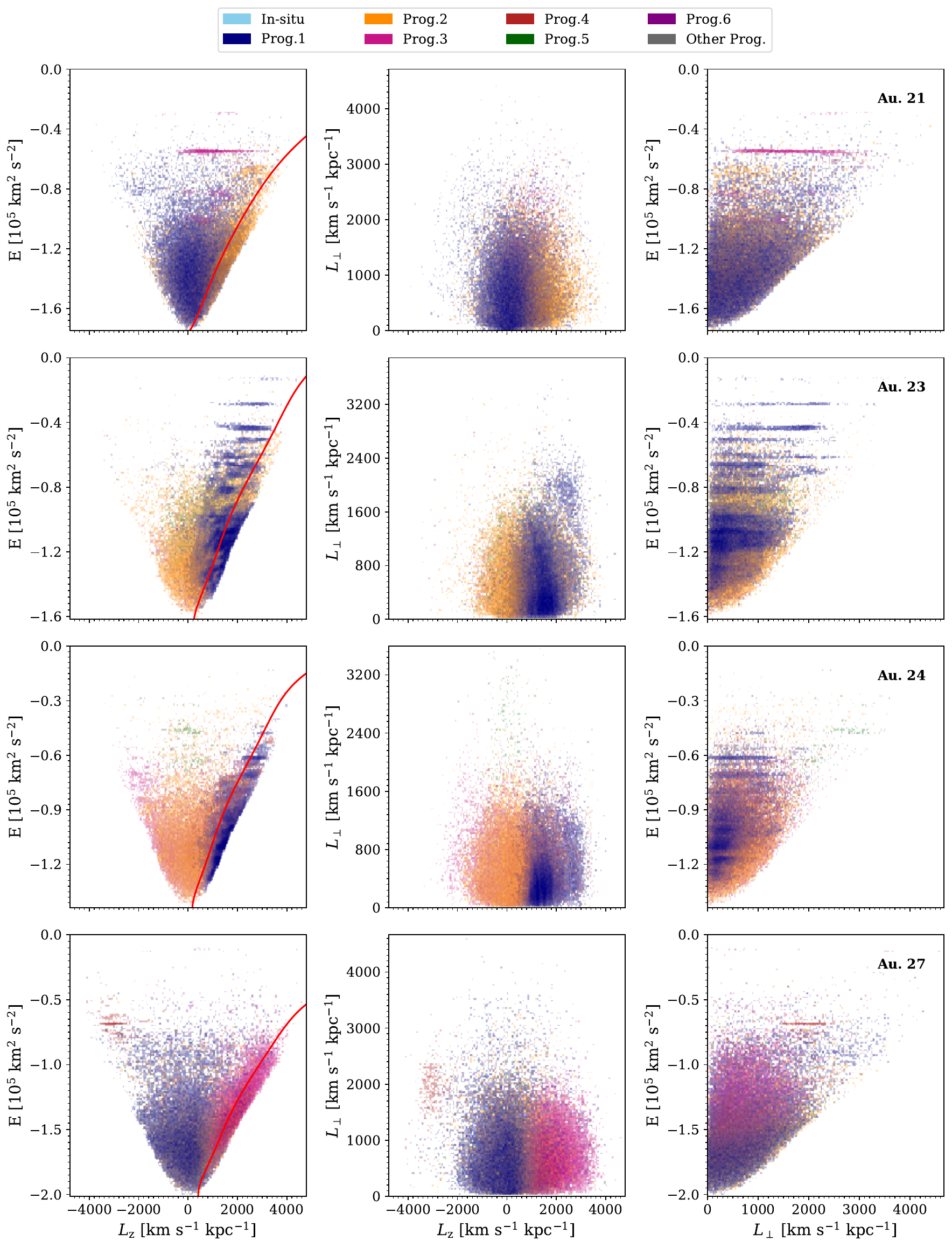}
   \caption{Same figure as Fig.~\ref{fig:dominance} but removing the in-situ component.} 
\label{fig:dominanceAccOnly}
\end{figure*}
\subsection{Properties of the most massive accreted galaxies}
In the Auriga simulations, the stellar particles of the simulated galaxies are categorized either as formed {\it in-situ}, accreted or found in existing sub-haloes ({\sc AccretedFlag $-1$, 0, 1}, respectively). Following the definition in \citet{monachesi_2019}, and used in several works based on this suite of simulations \citep[e.g.][]{fattahi_2019,thomas_2021}, the accreted stellar component contains the stellar particles that are gravitationally bound to the host galaxy at $z = 0$ but that were 
born in a satellite galaxy, i.e. particles that were bound to a satellite
galaxy in the snapshot when they were first identified. The {\it in-situ} component comprises the stellar particles that were bound to the host at ‘birth time’. Therefore, with this definition, the stars formed in the host from gas accreted from a satellite galaxy are flagged as {\it in situ}. Note here that we do not take into account the particles that are still bound to a surviving satellite galaxy, i.e. we only consider particles with {\sc AccretedFlag} $=-1$ to trace the {\it in-situ} component and 0 for the accreted one.

Among the accreted systems, in this work each progenitor is treated individually. However, to facilitate the analysis and the visualisation of the results, we list the individual properties only of the six most massive progenitors within each simulated galaxy, and group together the smaller ones. However, as we will show later, this distinction is not critical, as smaller progenitors do not significantly contribute to the formation of notable clusters in the Solar vicinity, with the exception of one cluster out of 111 in Au.23. The mass ranking is performed according to a progenitor' mass contribution to the {\it overall} stellar mass budget of the main galaxy. Fig~\ref{fig:massZprog} presents for each MW-like galaxy studied here, the current ($z=0$) stellar mass and the redshift of accretion of the six most massive progenitors, as well as the stellar mass of the in-situ component, in the entire galaxy (left) and in the Solar vicinity region (right). These parameters are also listed in Table~\ref{tab:param}, along with the ratio of  total and stellar mass between these accreted galaxies and the main galaxy at the infall time, i.e. the first moment when these satellite crossed the Virial radius. We note that the ranking of a given progenitor can change depending on the volume being considered, due to the way debris are shed as a consequence of the intricate interplay between a progenitor's mass, its epoch of accretion, and its orbital properties. It is important to point out that the progenitor ranking is performed using only the particle labelled as accreted, and not those still bound to a surviving progenitor. However, for the majority of progenitors, this does not impact the ranking, as most progenitors are already mostly, if not completely, disrupted. The only exception is Prog.~4 of Au.~21, for which 84\% of its stellar mass is still bound to a progenitor, despite having passed twice to its pericenter (at 40 kpc) in the last 3.5 Gyr. Considering the surviving progenitor, this galaxy will be the most significant accreted galaxy in terms of stellar mass. The total stellar mass of the galaxies given in Table~\ref{tab:param} differs slightly from the values quoted in \citet{grand_2024} due to the different methods used to account for the particles of the main galaxy. However, this small change does not impact our analysis.

As it can be seen, there is a range of accretion redshifts, with a few instances of recent accretions, $z<1$. This, combined with the orbital 
history and mass of each given progenitor, results in a diversity of morphologies in the $z=0$ spatial distribution of accreted stellar particles (an example is illustrated in Fig.~\ref{fig:St_M_D}). In particular, the $z<1$  accretions are in general found to retain spatial coherence at present-day, either as stellar streams or as disc-like structures, although Prog.~1 and~2 of Au.~21 are counter-examples. Therefore, we visually classify the debris of the six most massive progenitors in stream-like or disc-like structures, or as spatially-mixed if no structure is clearly identifiable. The state of each progenitor is indicated in Table~\ref{tab:param}.

Perhaps surprisingly, we observed that the debris of some of the most massive progenitors end up in disc-like structures, aligned with the host' disc and prograde, although these structures are slightly thicker. In the MW, the few clearly prograde structures identified up to-date do not have a disc-like morphology and tend to have polar orbits instead \citep[i.e. Helmi stream, Cetus-Palca, Sgr,][]{koppelman_2019,thomas_2022,ruiz-lara_2022}. The only notable exceptions are the Aleph and Nyx structure both prograde, with Aleph having a low eccentricity \citep{naidu_2020}, while Nyx have higher eccentricities than the thick disc \citep{necib_2020}. However, their chemo-dynamical properties tend to indicate that this structures does not emerge from an accreted galaxy, but rather originated from the Galactic disc itself \citep{zucker_2021,horta_2023}, like the Monoceros ring, A13, Triangulum-Andromeda stream  \citep{martin_2007,sharma_2010,gomez_2016,laporte_2018b}. Among the possibilities to explain that all these disc-like structures are prograde w.r.t to the host' disc is that their progenitor was massive enough to provide a torque on the disc, either by direct tidal stripping \citep{villalobos_2008,yurin_2015,gomez_2017}, or by the generation of asymmetric features in the dark matter halo of the host galaxy \citep{debattista_2013,gomez_2016,garavitocamargo_2020}, which ultimately led to align the spin axis of the disc with the orbital angular momentum of the satellite. This scenario is in agreement with the conclusion drawn by \citet{gomez_2017}, who observed the presence of a prograde disc-like accreted structure (referred to as {\it ex-situ} disc in their paper) in one-third of the simulated galaxy they studied. In the first case, the induced torque may primarily result from the infall of gas from the accreted galaxy, which subsequently reforms a disc in the host galaxy aligned with the axis of the merger. This scenario is particularly plausible given that most of these massive progenitors are gas-rich and are typically accreted during the early phase of the main galaxy. However, this process is not systematic, as illustrated by Au.23 and Au.24, where massive prograde accretions occurred relatively recently when the disc of the main galaxy is already well in place \citep{gomez_2017}. The detailed analysis of the precise mechanisms driving this effect is beyond the scope of this paper and will be addressed in a future study.

To get an intuition of whether given regions of the integral-of-motion space are dominated by particles proceeding from an individual progenitor or by multiple progenitors, we show the distribution of these six accreted massive galaxies in the total energy ($E$), vertical angular momentum ($L_z$), and perpendicular angular momentum ($L_\perp$) space in the form of ``dominance diagrams''. In practise, we colour-code bins in these quantities with a hue scale that reflects whether that bin is mainly populated by particles from one progenitor or from a mix; for example, a blue corresponding exactly to the blue used for Progenitor 1 in the legend implies that a given bin is entirely populated of particles originating from Progenitor 1\footnote{However due to the limitation of the eye perception, we can put a saturation of a unique colour if a bin is composed of at least 80\% by particles from a given progenitor, in particular if the second main contributor is colour-coded by a colour with a near hue.}, while the colours blends area show where there is a mix of progenitors. Fig.~\ref{fig:dominance} shows the case when both {\it in-situ} and accreted particles are considered, while Fig.~\ref{fig:dominanceAccOnly} shows the case when only accreted particles are considered. As it has been noted in several previous works \citep{jean-baptiste_2017,orkney_2023,horta_2024}, there are ample regions of the parameters space in which the debris from multiple progenitors overlap; at the same time, one accreted galaxy can give rise to multiple clumps in the integral-of-motion space.

When considering only accreted particles, it is possible to distinguish several areas in which a given galaxy is the dominant progenitor (for example, in the prograde region of Au.~23 and 24, or for some clumps at high energy in Au.~21 and 27). However,  not all massive accreted galaxies do have areas in this space where they are clearly dominant; for example, in Au.~21 and 27 one can distinguish three main colours rather than six, in Au.~23 two colours, and in the best case of Au.~24 four colours are ``dominant''. 

When including the in-situ component, everything becomes more interfused and dominated by in-situ particles, and the regions in which the dominance of a given accreted progenitor is visible are further reduced. This already gives a hint to the fact that most of the clumps that we will detect will be actually dominated by in-situ stars even when applying a cut in velocity to select halo stars, unless one finds an alternative way to remove the clearly in-situ component \citep[for example on the basis of some elemental ratio][]{hawkins_2015,das_2020,horta_2021,belokurov_2022}.

\section{Methodology} \label{sec:method}

To identify the signatures of accreted galaxies, we follow the methodology by \citetalias{lovdal_2022}. This involves searching for over-densities that could be caused by merger debris in the IoM space using a single linkage-based clustering algorithm. The significance of these clusters is quantified by comparing them to artificially created smoothed haloes.

The hypothesis behind this approach is that the debris of an accreted galaxy remain clustered in IoM space even several Gyr after the complete dissolution of the progenitor \citep{helmi_1999}. It is also implicitly assumed that the gravitational potential of the host galaxy can be well approximated by a time-independent axisymmetric potential, since the integrals-of-motions being used are the total energy ($E$), the vertical angular momentum ($L_z$), and the perpendicular angular momentum ($L_\perp$). All of these assumptions are, in reality, somewhat broken. For instance, $L_\perp$ is not fully conserved in an axisymmetric potential. Nonetheless, it is often used for searching signatures of accreted galaxies, as it is expected that stars originated from the same progenitor remains clustered for several Gyr in this parameter space \citep[e.g.][]{helmi_1999,williams_2011,jean-baptiste_2017,koppelman_2019}. Even regarding the other parameters, it has been shown that the quantities are not always conserved during and after the accretion, in particular for the most massive galaxies, as their energy and angular momentum decrease rapidly due to dynamical friction, which can result in several local overdensities at different energy/angular momentum level \citep[e.g.][]{jean-baptiste_2017,grand_2019,amarante_2022,khoperskov_2023}. Another source of limitation of these assumptions is that the Galactic gravitational potential is not axisymmetric, due to the presence of the bar, spiral arms, or large scale perturbations that affect the dynamics of disrupted structures \citep[see][for the impact of non-axisymetries on stellar streams]{pearson_2017,vasiliev_2021,thomas_2023}, and it is varying in time \citep[e.g.][]{buist_2015,koppelman_2021}.

The algorithm of \citetalias{lovdal_2022} to detect substructures in the IoM space is succinctly outlined below, but we refer the reader to their sections~3 and 4 for a more detailed description.

The first step is based on the single linkage method \citep{everitt_2011} to identify potential clusters in the IoM. This hierarchical clustering procedure incrementally connects the two closest groups of stars (or two individual stars) that are not yet connected according to a given metric. The metric chosen by \citetalias{lovdal_2022} was the Euclidean distance between two (groups of) stars in the {\it scaled} IoM space. Indeed, to ensure that each IoM parameter ($E$, $L_z$, $L_\perp$) is equally important when searching for clusters, each of them is rescaled such that the values of the distribution are in the range [$-$1,1]. 

In the second step of the algorithm, the significance of each potential cluster ($C_i$) found in the previous step is computed by comparing the number of stars belonging to the cluster, $N_{C_i}$, to the average number of stars $ \langle N^{art}_{C_i}\rangle$ measured in the same region of the IoM space in a set of $\mathcal{N}$ artificial haloes, whose generation process is described in the Sect.~\ref{sec:artHalo}. As done by \citetalias{lovdal_2022}, we only investigated candidate clusters with at least ten members, since smaller clusters are not significant assuming Poissonian statistics. It is important to acknowledge that this minimum threshold may restrict our ability to detect debris from low-mass accreted galaxies with stellar masses $\sim 10^6$ M$_\odot$, since they are composed of $\sim 100$ stellar particles in the simulations used here.  Following \citetalias{lovdal_2022}, the area covered by of each potential cluster is assumed to be elliptical, with the axis lengths equal to $a_i=2.83 \sqrt{\lambda_i}$, where $\lambda_i$ are the eigenvalues obtained using a principal component analysis \citep[PCA,][]{pearson_1901,hotelling_1933} on stars composing the potential cluster.  As \citetalias{lovdal_2022}, the clusters below a significance threshold of $3\sigma$ are discarded, such that only the clusters where $N_{C_i} - \langle N^{art}_{C_i}\rangle \geq 3\sigma_i$ are conserved. Here $\sigma_i=\sqrt{N_{C_i}+(\sigma^{art}_{C_i})^2}$, where $\sigma^{art}_{C_i}$ is the standard deviation of the number of stars detected over the set of $\mathcal{N}$ artificial haloes\footnote{It is worth noting that, since our goal is to measure how much the observed number of stars (or particles) in a region of the IoM space deviates from the average, accounting for the scatter due to different realizations of artificial background halos, adding Poissonian noise may not be the most optimal approach. This could potentially lead to an under-detection of small systems. A comparison between these two metrics would be an interesting avenue for future work.}.  
Some identified clusters with at least a significance of $3\sigma$ overlapped to each other in the IOM space and are linked together by the single linkage method. Therefore, by making the assumption that the significance increases by adding stars of the same structure and decreases by adding noise, one can select the final exclusive clusters by searching the location of the maximum significance of connected significant clusters (more than $3\sigma$). In practice, this is done by exploring the merger tree of the significant clusters obtained by the single linkage method, ordered by descending significance \citepalias[see Section~4.1.2 of][]{lovdal_2022}.

In \citetalias{lovdal_2022} and \citetalias{dodd_2023}, the number of stars in each cluster is refined by retaining only those with a Mahalanobis distance from the cluster centre of less than 2.13. This approach preserves 80\% of the original cluster members, assuming the stars in a cluster follow a multivariate Gaussian distribution in the IoM. Furthermore, they include extra members by adding all stars located within that Mahalanobis distance range, even if they were not part of the original clusters according to the single linkage method. However, we found that this revised selection has a minimum impact on our results (see discussion about it in Sect.~\ref{sec:recovrate}). Therefore, decided to keep the original cluster populations, without adding extra members. 

The steps outlined above are solely based on the stars' IoM properties. In a companion article, \citetalias{ruiz-lara_2022} also investigated whether incorporating additional information about the stellar population properties—such as the average age and metallicity distribution function—could provide further insight into the reality of these features. In Sects.~\ref{sec:chemage} and \ref{sec:acconly}, we will explore this approach as well.

\begin{figure}
\centering
  \includegraphics[angle=0,clip,width=9cm]{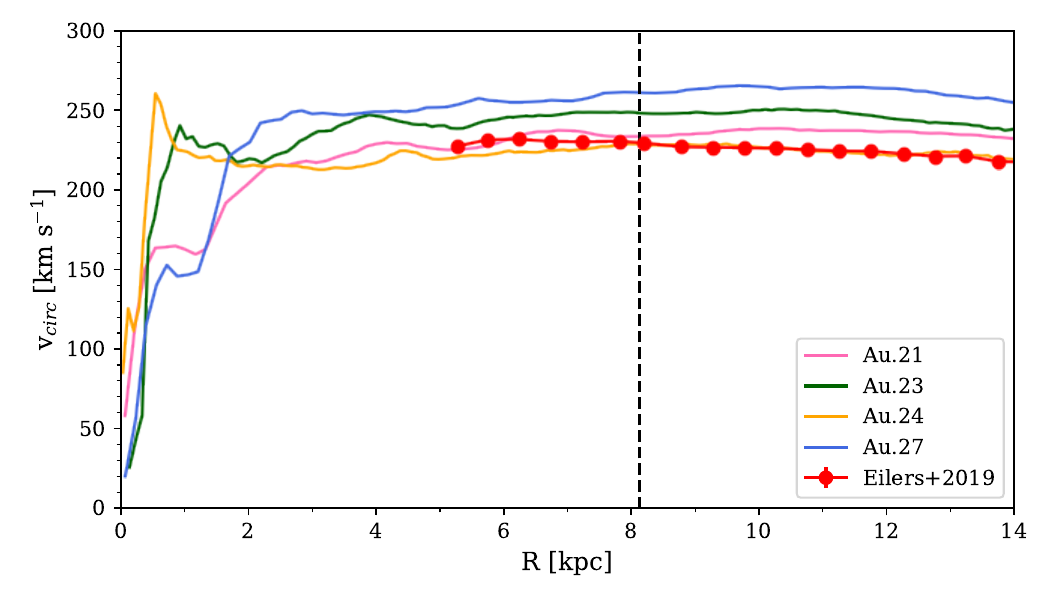}
   \caption{Circular velocity curve of the four MW-like galaxies studied here. We note that the radii have been rescaled as explained in Section \ref{sec:overview}. The Solar radius ($R_\odot=8.2$~kpc) is indicated by the black vertical line. For comparison, we show the rotation curve of the Milky Way measured by \citet{eilers_2019}.} 
\label{fig:vcirc}
\end{figure}

\begin{figure}
\centering
  \includegraphics[angle=0,clip,width=9cm]{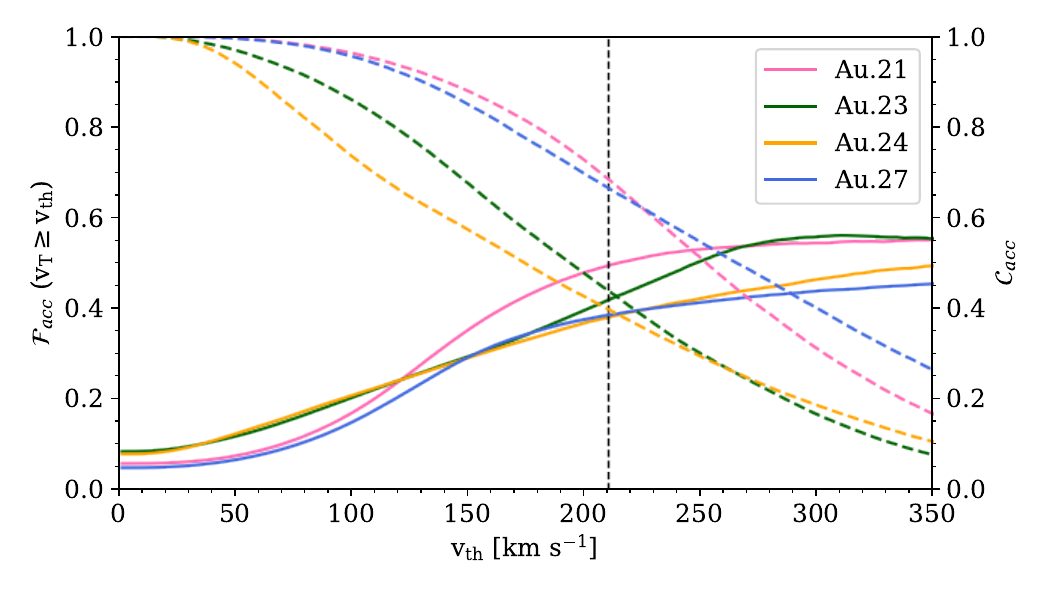}
   \caption{Fraction of accreted stellar particles ($\mathcal{F}_{acc}$, solid lines) in the Solar vicinity with velocities above a given threshold velocity ($V_\mathrm{th}$), for the four simulated galaxies. The dashed lines show the evolution of the completeness of accreted stellar particles ($\mathcal{C}_{acc}$) as a function of the threshold velocity. The vertical dashed line indicates a velocity threshold of 210 km s$^{-1}$, used to kinematically define the halo in the Solar vicinity.} 
\label{fig:vt_cut}
\end{figure}

\subsection{Adapting the methodology to the analysis of the Auriga simulations}

\subsubsection{Selection of the halo sample in the Solar vicinity} \label{sec:haloSun}
Akin to what is done in several works to identify the local stellar halo of the MW in data \citep{koppelman_2018,koppelman_2019,lovdal_2022,dodd_2023}, we select particles with halo-like kinematics requiring them to have a total velocity with respect to the local standard of rest (LSR)  ($\mathrm{V_T}\equiv |{\bf V} - {\bf V_\mathrm{LSR}}|$) above a given threshold velocity ($\mathrm{V_{th}}$), chosen to remove the large majority of the particles of the galactic disc.

In the simulations, ${\bf V_\mathrm{LSR}}$ is equal to the circular velocity at the solar radius $V_{circ} (R_\sun)$. For each simulated galaxy, we obtain the circular velocity curve (presented in Fig.~\ref{fig:vcirc}) from young ($\leq$ 1~Gyr old) {\it in-situ} stellar particles located within 1~kpc of the Galactic plane. The circular velocity at the Solar radius is V$_{circ} (R_\sun)= 229.5$~km~s$^{-1}$ for simulation Au.~21, $261.4$~km~s$^{-1}$ for Au.~23, $207.2$~km~s$^{-1}$ for Au.~24 and $261.1$~km~s$^{-1}$ for Au.~27.

Ideally, one would like to fix the threshold velocity ($\mathrm{V_{th}}$) used to select the stellar halo such that the majority of the particles with $\mathrm{V_T}\geq \mathrm{V_{th}}$ originated from accreted galaxies. However, as visible in Fig.~\ref{fig:vt_cut}, this is not possible in all simulations, as the fraction of accreted stars remains at most of 45-55\% even for $\mathrm{V_{th}} = 350$~km~s$^{-1}$, at which point the completeness would drop to 10-25\%.  Therefore, we decided to use the same threshold velocity than \cite{koppelman_2018,koppelman_2019} and \citetalias{lovdal_2022}, i.e. $\mathrm{V_{th}}=210$~km~s$^{-1}$, since it yields a fraction of accreted stellar particles not too dissimilar than when adopting larger thresholds, but for a much higher level of completeness.

\begin{table}
\centering
\caption{Boundary of each IoM parameter used to rescale their distribution  into the range [-1,1] for the dynamically selected halo. }
\label{tab:renorm}
\begin{tabular}{lccc}
Galaxy & $E$ [km$^2$~s$^{-2}$] & $L_z$ [kpc~km~s$^{-1}$] & $L_\perp$ [kpc~km~s$^{-1}$]\\
\hline
Au.21 & $[-1.75\ 10^5; 0]$ & $[-4416; 4567]$ & $[0; 4287]$ \\
Au.23 & $[-1.61\ 10^5; 0]$ & $[-4240; 4437]$ & $[0; 3802]$ \\
Au.24 & $[-1.44\ 10^5; 0]$ & $[-3891; 3962]$ & $[0; 3609]$ \\
Au.27 & $[-2.00\ 10^5; 0]$ & $[-4995; 5302]$ & $[0; 4437]$ \\
\hline\\
\end{tabular}
\end{table}

\begin{figure*}
\centering
  \includegraphics[angle=0,clip,width=17cm]{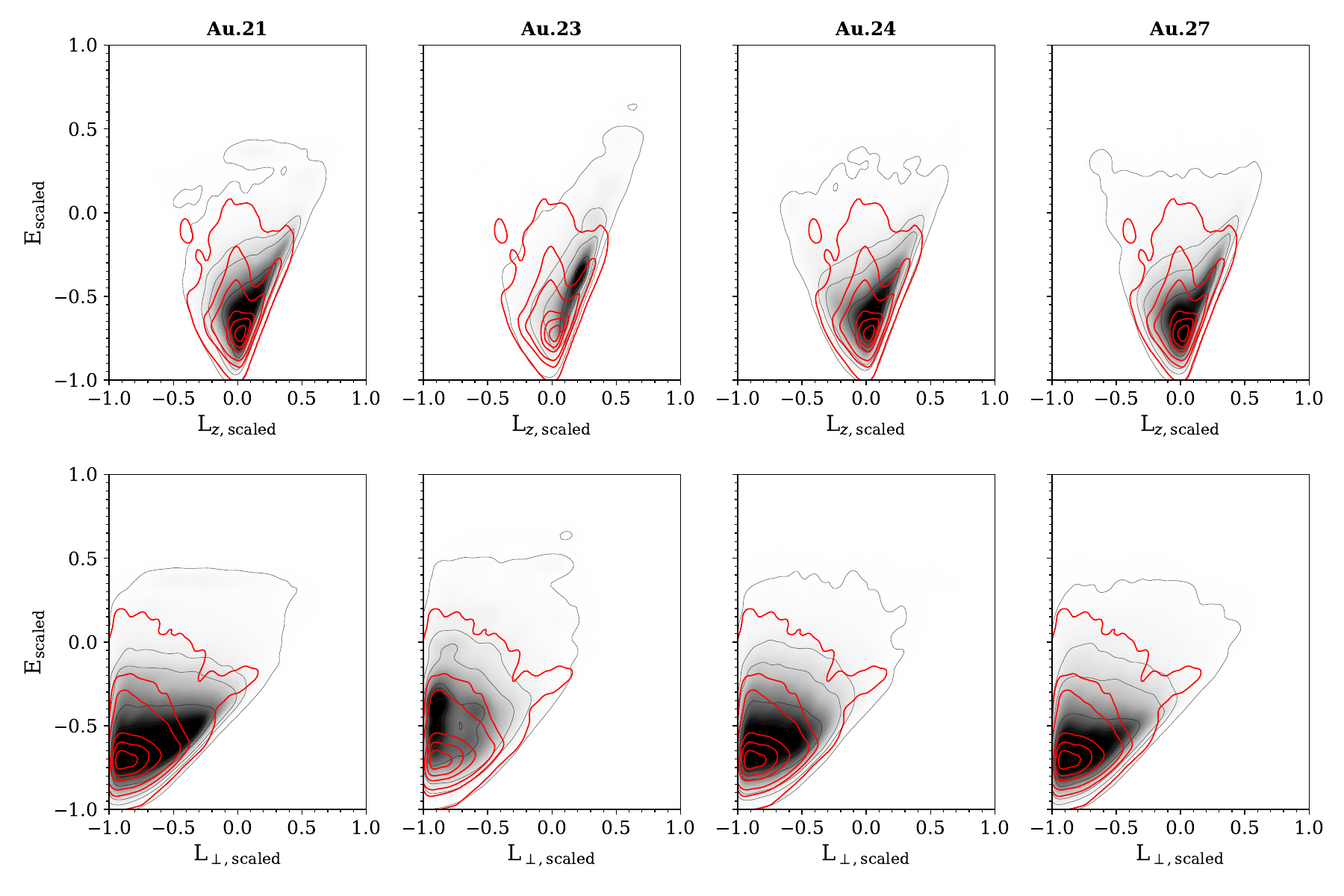}
   \caption{Distribution of the rescaled IoM parameters of the kinematically-selected local halo sample of the four simulated galaxies (black contours). As a comparison, in each panel, the distribution of the local MW halo sample by \citetalias{lovdal_2022} is shown by the red contours. In both cases the iso-density contours are plotted at the 1, 10, 20, 50 and 90\% of the maximum density.} 
\label{fig:renorm_param}
\end{figure*}

\subsubsection{Calculation of IoM and renormalization in IoM space} \label{sec:description}

We compute the total energy of each particle using directly the gravitation potential of the simulated system.

In our case the factors used to re-normalize the distribution of $E$, $L_z$, and $L_\perp$ are different for each {\it simulated} galaxy, given that they cover a different range of the parameter space. The values used to renormalize the IoM space for each galaxy, listed in Table.~\ref{tab:renorm}, correspond to the approximate minimum and maximum range spanned by each of these parameters for the stellar particles in the Solar vicinity. The renormalized parameters space of the local halo sample of each simulated galaxy is compared to the one of the \citetalias{lovdal_2022} sample in Fig.~\ref{fig:renorm_param}. In this figure, we can see that, with the notable exception of Au.~23, which has a clear over-density of particles with high vertical angular momentum, all the simulated systems have a distribution in the renormalized IOM space similar to that observed in the MW Solar neighbourhood. However, we can also see that none of them present a structure similar to the Gaia-Enceladus Sausage in the energy-vertical angular momentum plane, which in the data is clearly visible by the almost vertical overdensity around $(\mathrm{L}_{z,scaled},~\mathrm{E}_{scaled})=(0,0)$. This is very interesting, as \citet{fattahi_2019} found that Au.~24 and Au.~27 have a progenitor with current characteristic similar to Gaia-Enceladus-Sausage (GES). However, their criteria, based on the velocity anisotropy, and the region ($9<|z| < 15$~kpc) used to identify this progenitor, are different from the criteria we used and the region we study. \citet{orkney_2022} also identified a GES-like progenitor in Au.~24, which was chosen to have a redshift of accretion and a stellar mass similar to the estimation obtained for GES \citep[$z\simeq 2 - 1.5$, $M_* \sim 10^8-10^9$ M$\odot$,][]{helmi_2018,mackereth_2019,gallart_2019,naidu_2022,lane_2023}. Their GES-like progenitor corresponds to our Prog.~2 of Au.~24, and their Kraken-like progenitor corresponds to Prog.~3. In their Fig.~3, we can see that the GES structure presents a vertical profile centred on $\mathrm{L}\mathrm{z}=0$ km~s$^{-1}$~kpc$^{-1}$. However, the region they are studying is larger than the volume of the Solar vicinity we are using in this study, which naturally tends to increase the energy it spans, and the scattering in $\mathrm{L}\mathrm{z}$ is significantly larger than the actual scatter of GES.

\subsubsection{Generation of the artificial haloes} \label{sec:artHalo}

As mentioned in a previous section, the significance of each cluster is computed by comparing the number of particles inside an IoM region to that in a set of $\mathcal{N}$ artificial halo. Similarly to \citetalias{lovdal_2022}, the artificial haloes are made by scrambling two of the three the velocity components of the original dataset (i.e. in practice $v_y$ and $v_z$). By doing so, these artificial haloes are smoother than the original dataset, since the structures present in the integral-of-motion space are washed out \citep{helmi_2017}.
However, by scrambling the velocities, this method tends to increase the kinetic energy of individual stars, and thus their total energy, such that some stars with high potential energy might finish having a positive total energy in these artificial haloes, i.e. to not be bound to the galaxy anymore. 

To palliate this problem, \citetalias{lovdal_2022} generated the artificial haloes by scrambling the velocity of an extended halo sample, selected using a less restrictive kinematic criterion of $\mathrm{V_{th,art}}=180$~km~s$^{-1}$, instead of $\mathrm{V_{th}}=210$~km~s$^{-1}$ for the original halo sample. The artificial halo is then obtained by sub-selecting $N$ stars with $\mathrm{V_T} \geq 210$~km~s$^{-1}$ from this extended halo sample, where $N$ is the number of stars present in the original halo dataset. 

In our case, because the distribution in the IoM space is very different across the simulated galaxies to another, it is not possible to use a unique value of $\mathrm{V_{th,art}}$. Our experiments revealed that using a common value of $\mathrm{V_{th,art}}$ for all the simulation led to notable discrepancies between the artificial haloes and their corresponding original counterparts. In particular, the artificial haloes consistently exhibited higher average vertical angular momentum than the original haloes.

To determine the optimal $\mathrm{V_{th,art}}$ for each case, we conducted a systematic analysis where we calculated the average values of IoM parameters ($E$, $Lz$, and $L_\perp$) across 10 artificial haloes for various $\mathrm{V_{th,art}}$ values, ranging from 50 to 210~km~s$^{-1}$ with an increment of 5~km~s$^{-1}$. The correct $\mathrm{V_{th,art}}$ value was identified as the one at which the normalized Euclidean distance ($D$) between the means of IoM parameters in the original halo dataset and those in the set of 10 artificial haloes was minimized. Mathematically, this is expressed as:
\begin{equation}
    D=\sqrt{\sum_{X=E,~L_z,~L_\perp} \left( \frac{\langle X_{art}\rangle-\langle X_{ori}\rangle}{\sigma_{X,ori} }\right)^2 },
\end{equation}
where $\sigma_{X,ori}$ is the standard dispersion of each IoM parameters in the original halo.

\begin{figure*}
\centering
  \includegraphics[angle=0,clip,viewport=0 130 1025 754,width=17cm]{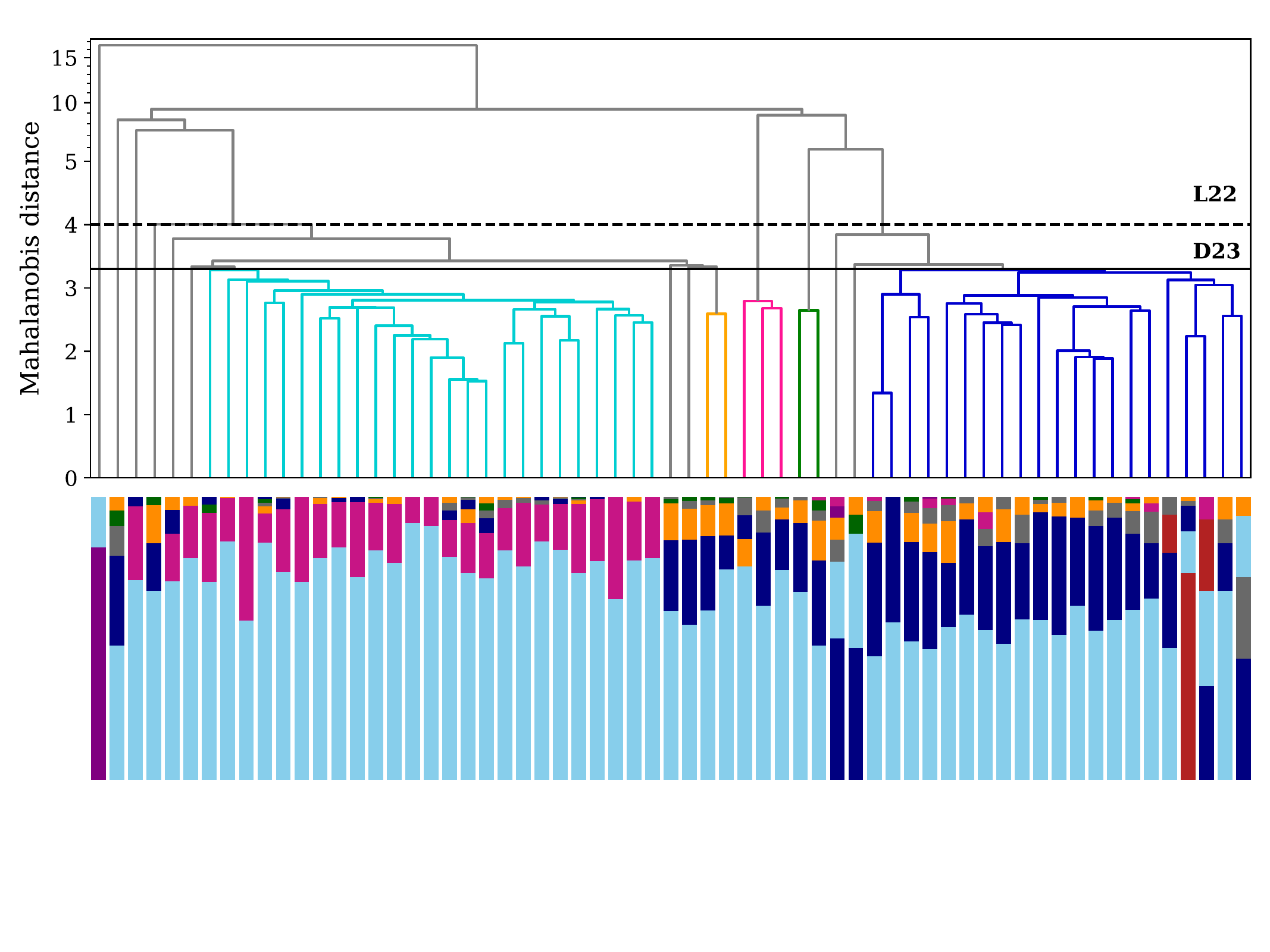}
   \caption{A dendrogram similar to Fig. 10 of \citetalias{lovdal_2022} shows the relationship between the significant clusters detected by the single linkage algorithm according to their Mahalanobis distance in the IoM space, with the same cut-off threshold as used by \citetalias{dodd_2023} (indicated by the black line). For reference, the cut-off threshold used by \citetalias{lovdal_2022} is shown by the dashed line. The vertical bars at the bottom of the figure indicate the relative contribution of the different progenitors to each significant cluster, using the same colour scheme as in Fig.~\ref{fig:dominance}. Note that the colours used to indicate the different groups of clusters linked together in the dendrogram are not related to the progenitors.} 
\label{fig:dendrogram}
\end{figure*}

\section{Analysis and results} \label{sec:results}

We have applied the methodology described in Sect.~\ref{sec:method} to the four Auriga haloes previously discussed. In the remainder of the article, we will use Au.~27 to exemplify some of the results because this halo presents the accretion history that is the most similar to the one currently estimated for the MW, i.e. with the two most important accretions occurring at $z\simeq 1-2$ \citep[see][and references therein for a recent review on the vision of the MW history]{deason_2024}, similar to GES and Kraken, and another important accretion in the last 4-6~Gyr, with the debris of the progenitor that still forms a stellar stream at $z=0$, similar to the Sagittarius stream in the MW. The results and plots concerning the other haloes will be included as online material.

Table~\ref{tab:cl27} lists the statistically significant clumps found for Au.~27, along with their main properties (number of particles, significance, fraction of particles accreted, mean position in the IoM, dominant progenitor, contribution of the dominant progenitor to the cluster), as well as their suggested association in groups on the basis of a threshold in Mahalanobian distance in IoM space between the clusters (see Sect.~\ref{sec:GroupClumps}) and with the addition of the information on the metallicity, magnesium abundance, and age distribution of the stellar particles belonging to them. We find 63 clusters, with statistical significance between 3 and 17 and containing from a few dozens to up to more than 4000 particles, with a median of 58 particles per cluster. These values are similar to those encountered by \citetalias{lovdal_2022}. This indicates that the kinematically selected stellar halo of Au.~27 presents a similar level of lumpiness as observed in the MW. The fraction of stellar particles that are associated to clusters is $\sim 9.4\%$, similar to the values found by \citetalias{lovdal_2022} and \citetalias{dodd_2023} for the MW ($\sim 13\%$). In the other simulated galaxies, the number of significant clusters found ranges from 49 (Au.~24) to 111 (Au.~23), with a fraction of stars in clusters varying from 4\% (Au.~24) to 13.5\% (Au.~23), reflecting the different accretion histories that occur in each of them. 

We will now examine the general purity and recovery rate (completeness) both of the individual clumps and then of the groups themselves. 

\begin{figure*}
\centering
  \includegraphics[angle=0,clip,width=17cm]{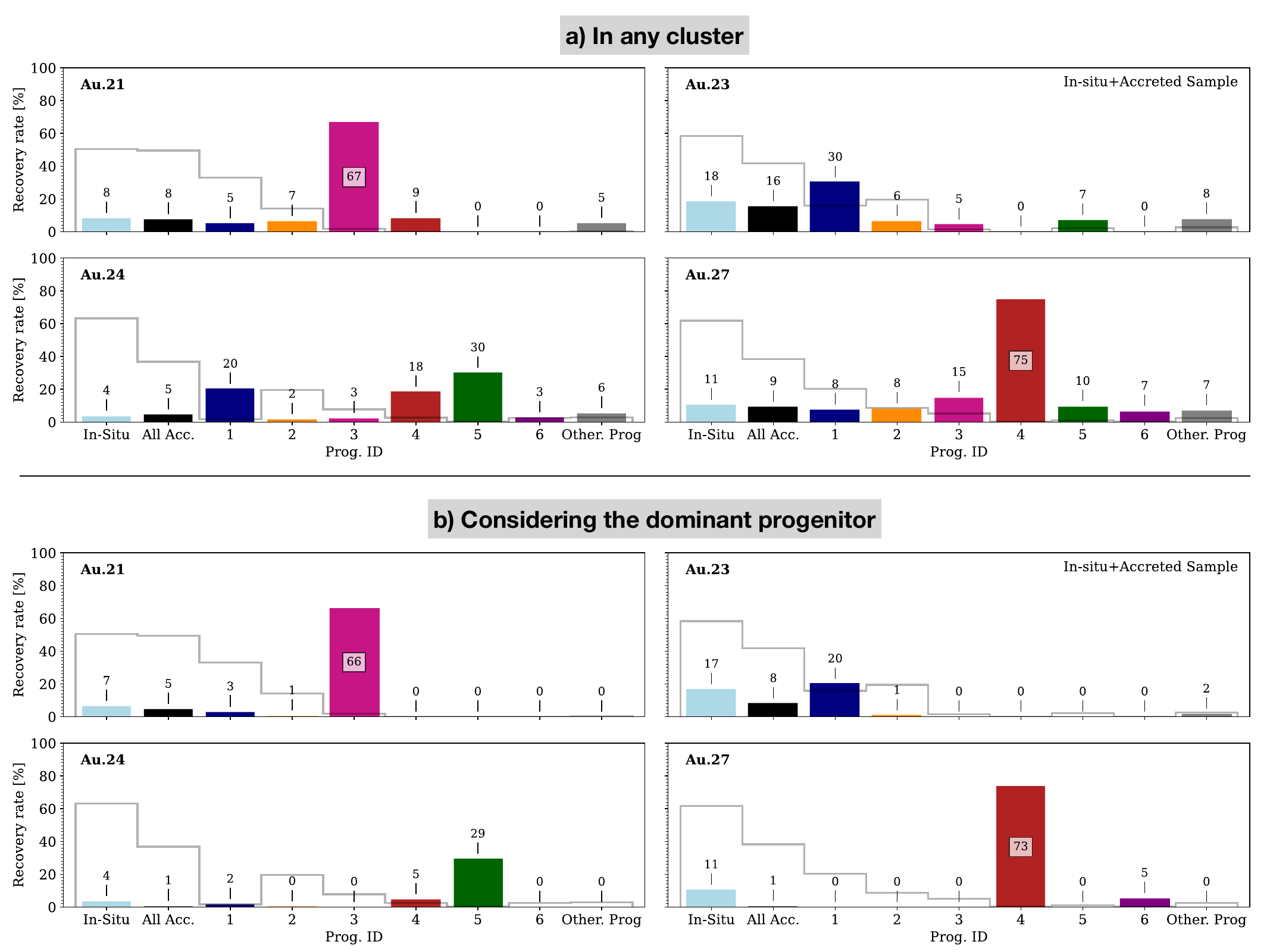}
   \caption{Percentage of stellar particles, separated by their birth environment, located in the kinematically selected halo in the Solar vicinity identified as being part of any significant cluster ($>3\sigma$) by the single-linkage method (upper panels). The lower panels show the similar quantity but considering only the particles that belong to clusters where their progenitor is the dominant contributor to a given clump. In all panels, the grey histograms indicate the contribution of each progenitor to the Solar vicinity halo. Note that the second bar represent the value of all the accreted stellar particles grouped together.} 
\label{fig:recov_ISacc}
\end{figure*}

\subsection{Recovery rate and purity of individual clumps} \label{sec:recovrate}

The bottom panel of Figure~\ref{fig:dendrogram} displays, for each of the individual statistically significant clumps, their composition in terms of birth environment/progenitor, with the length of the vertical bars being directly proportional to the percentage of particles belonging to that progenitor (the most dominant progenitor is found at the bottom of the bar). Across all haloes, the majority of clumps are heavily contaminated by the presence of in-situ particles, with the exact percentage varying from halo to halo. In Au.~27, and this is particularly the case for this simulation, in-situ particles are numerically dominant in all but a minority of clumps. This is a surprising result, as in-situ particles are not expected to form overdensities, unlike accreted particles. Some of these in-situ dominated clusters may result from the response of the Galactic disc  to past mergers \citep{gomez_2012,jean-baptiste_2017,laporte_2018,thomas_2019}, but this will  explain the presence of such clusters on prograde orbits. However, we found that between 20 and 40\% of clusters where in-situ particles make up more than 50\% of the population are located at L$_z<500$~km~s$^{-1}$~kpc$^{-1}$. 

Moreover, it is also rare to find clumps where all, or the great majority of, the particles originate from a single accreted progenitor, except for possible cases of in-situ particles formed from gas that was previously bound to accreted galaxies. For instance,  {in Au.~21(23), only 5(17) out of 54(111) clusters have more than 75\% of their particles originating from a single accreted progenitor, with just one such cluster in both Au.~24 and Au.~27. This number further drops to 2(5) clusters in Au.~21(23) and none in Au.~24 and Au.~27 when the threshold is increased to 90\%.}  

None of the simulations show a clear trend between the purity (i.e. the relative contribution of the dominant progenitor) and the significance of a cluster, regardless of the inclusion or not of in-situ particles. However, for significance higher than 7-8, we observe that the minimum purity of the clusters is of $\simeq 0.6$, which increases such that clusters with a significance >15 are pure at 90\%. This remains true even when the clusters are separated into different categories based on their population size. This is surprising, as one might have reasonably expected that the most significant clusters would be mainly composed of particles from a single progenitor since these clusters are much more populated compared to a smooth background. This lack of correlation between the purity and the significance indicates a limitation of the method, and might be linked to spurious over/under-densities from the generation of artificial backgrounds (see Sect.~\ref{sec:backgen}).

Fig.~\ref{fig:recov_ISacc} shows that the rate of recovering accreted particles as part of statistically significant clumps is in general below 10-15\%. This percentage drops to $<$5\% if we count only particles belonging to the dominant progenitor of a given clump\footnote{Note that the dominant progenitor does not always contribute to 50\% or more of the particles of a cluster.}. Exceptions to this behaviour are progenitor~3 in Au. 21, progenitor~1 in Au. 23, 1-4-5 in Au. 24 (only 5 for the case of the dominant particles), progenitor~4 in Au. 27; qualitatively speaking, these are essentially those progenitors that produce the tight clumpy features in energy visible in Fig.~\ref{fig:dominance}, i.e. typically those that have been recently accreted ($z<0.6;$ $<6$~Gyr ago), and are found either in a stream- or disk-like configuration and do contribute with a sufficient percentage of particles to the sample in the volume being analysed. In the four simulations, all the clusters with a significance higher than 6$\sigma$ not dominated by in-situ particles are dominated by particles originating by a galaxy accreted less than 6-7~Gyr ago. Interestingly, this relationship between the recovery rate and the age of accretion is independent of the total or stellar mass involved in the merger. Although all these recent accretion events are minor mergers, with a stellar mass ratio of less than $10:1$, their impact on the recovery rate appears negligible —provided the progenitors leave debris in the Solar Vicinity. This result is intriguing because, for a given accretion time, one might expect the debris of smaller progenitors to be more easily detectable, and thus having a higher recovery rate than most massive progenitors for which dynamical friction is more important, leading to a wider dispersion of debris in Integral of Motion (IoM) space \citep{jean-baptiste_2017} and, consequently, to a lower recovery rate.

\begin{figure*}
\centering
  \includegraphics[angle=0,clip,width=9.cm,viewport=0 0 530 405 ]{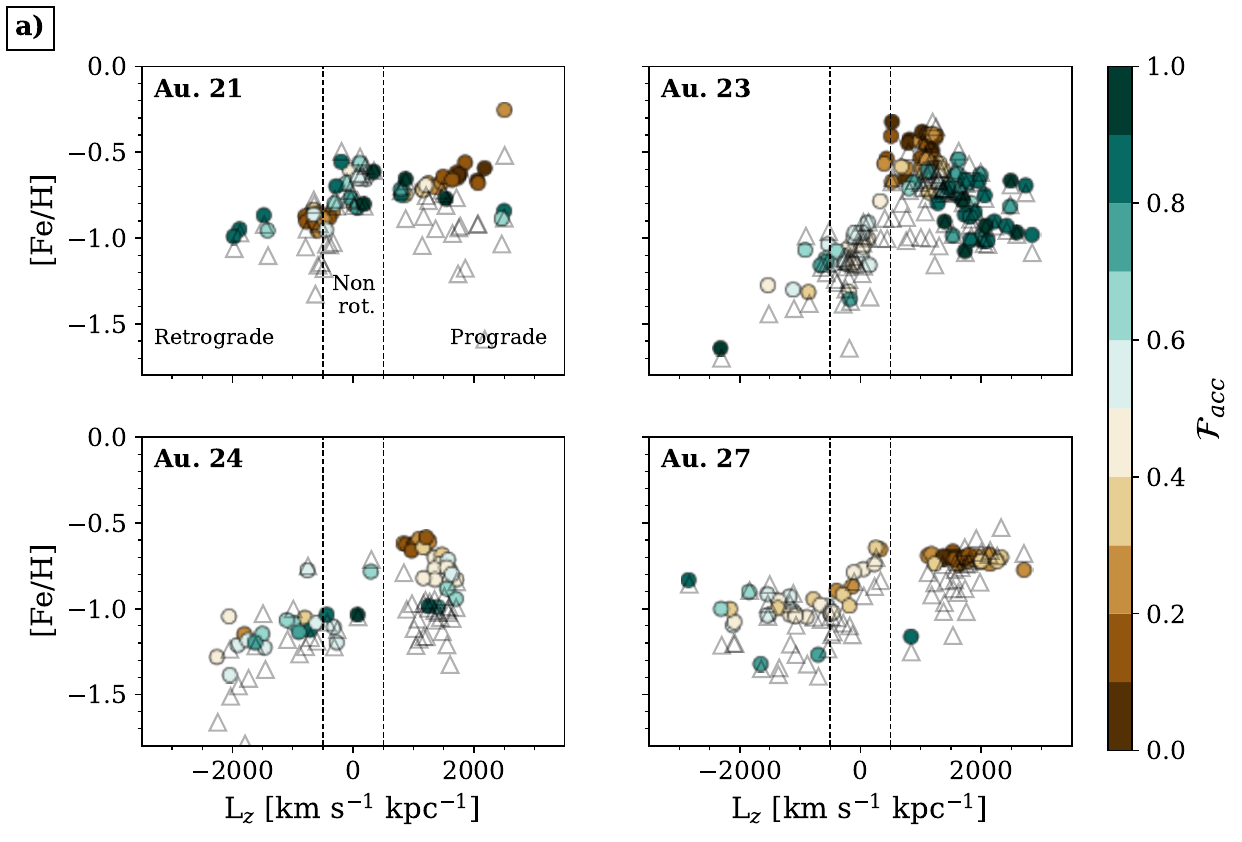}
  \vline
  \includegraphics[angle=0,clip,width=9.cm,viewport=-10 0 520 405 ]{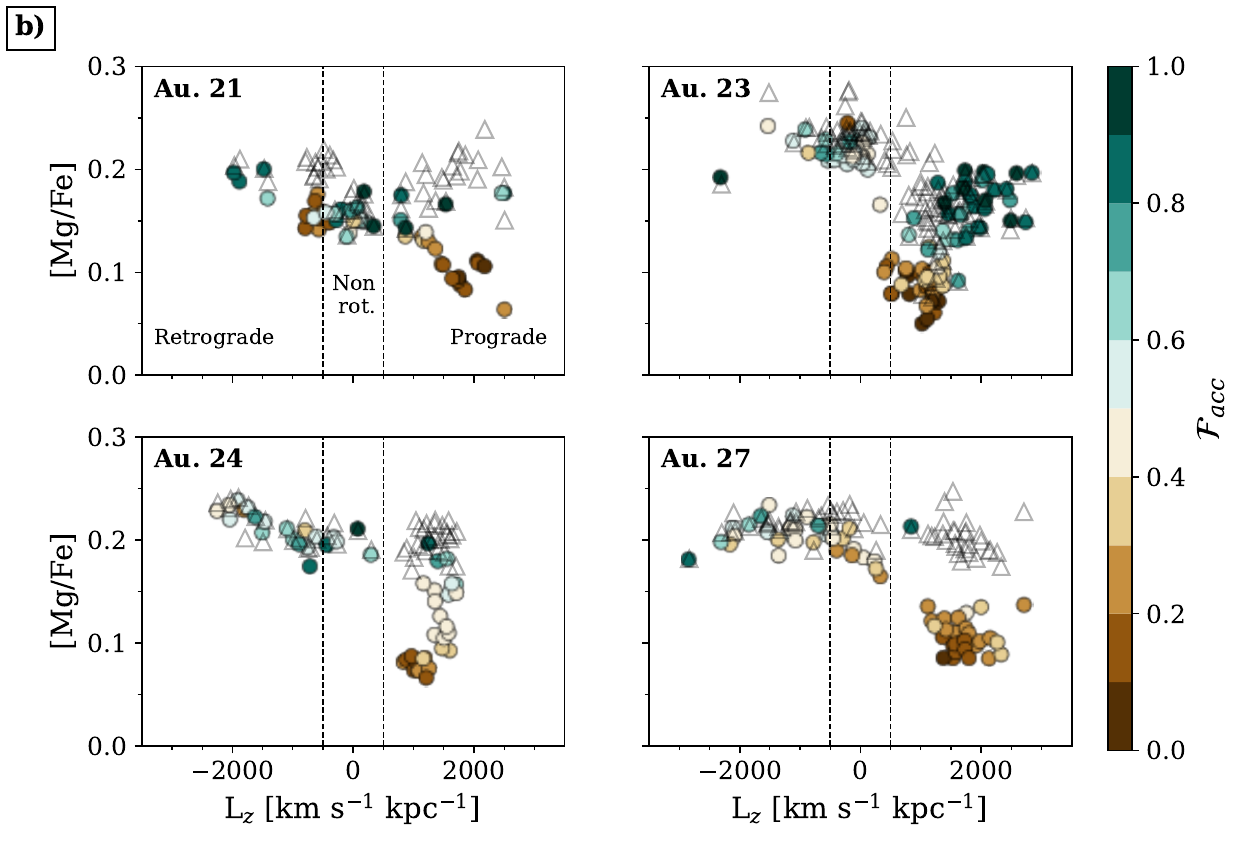}
  \includegraphics[angle=0,clip,width=9.cm,viewport=0 0 595 420 ]{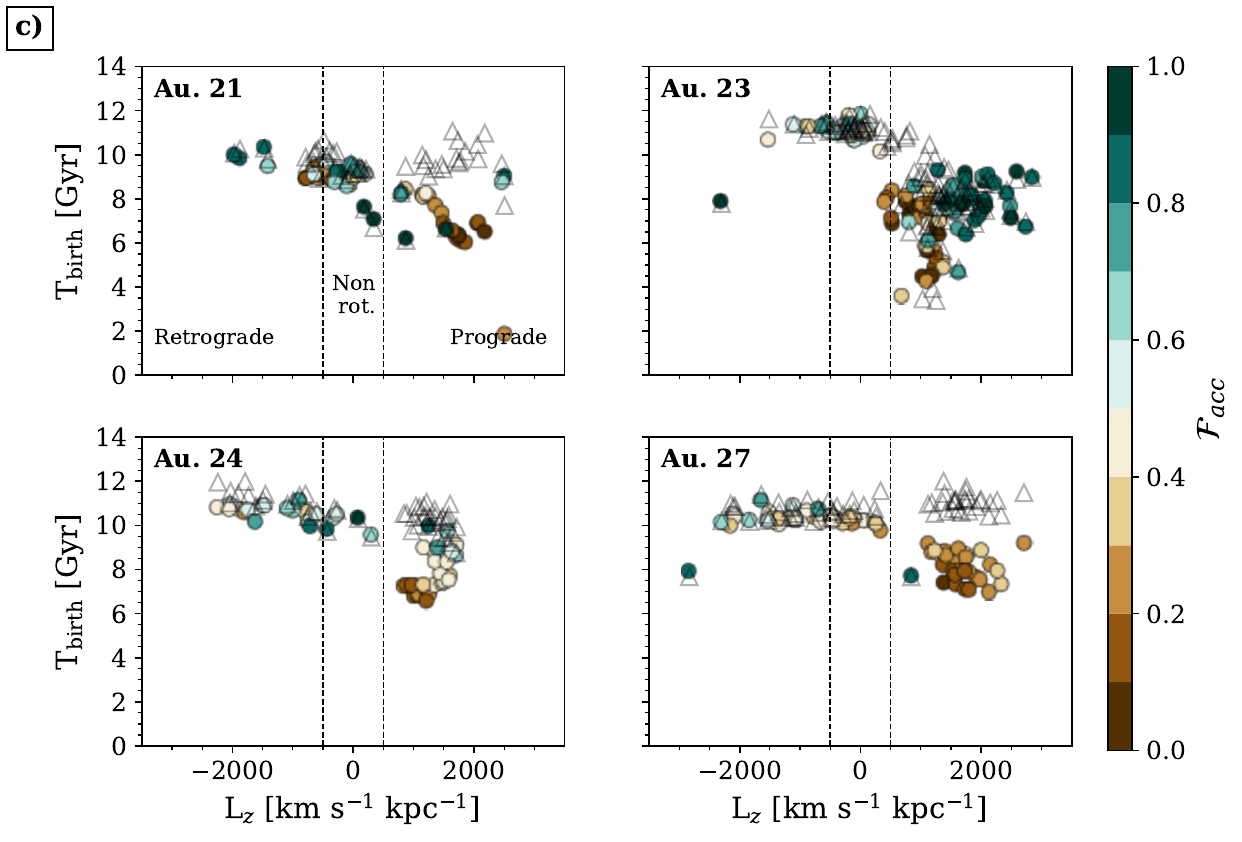}
   \caption{Median [Fe/H] (panels a), [Mg/Fe] (panels b) and stellar age (panels c) as a function of the mean $(\mathrm{L_z})$ of each cluster found in the kinematically selected halo of the Solar vicinity in the four simulated galaxies (circles). The circles are colour-coded by their fraction of accreted stellar particles. The triangles represent the same quantity, but with the in-situ particles removed. The vertical dashed lines located at $(\mathrm{L_z} \pm 500~\mathrm{km~s^{-1}~kpc^{-1}})$ indicate the separations between the retrograde, non-rotating, and prograde clusters. For the prograde clusters, the metallicity, but mostly the [Mg/Fe] and the stellar age separate well those dominated by in-situ and accreted particles. However, none of those parameters is able to disentangle the clusters dominated by in-situ to those dominated by accreted particles for the non-rotating and the retrograde clusters.} 
\label{fig:clLZage_feh_mg}
\end{figure*}

\begin{figure*}
\centering
  \includegraphics[angle=0,clip,width=18.5cm]{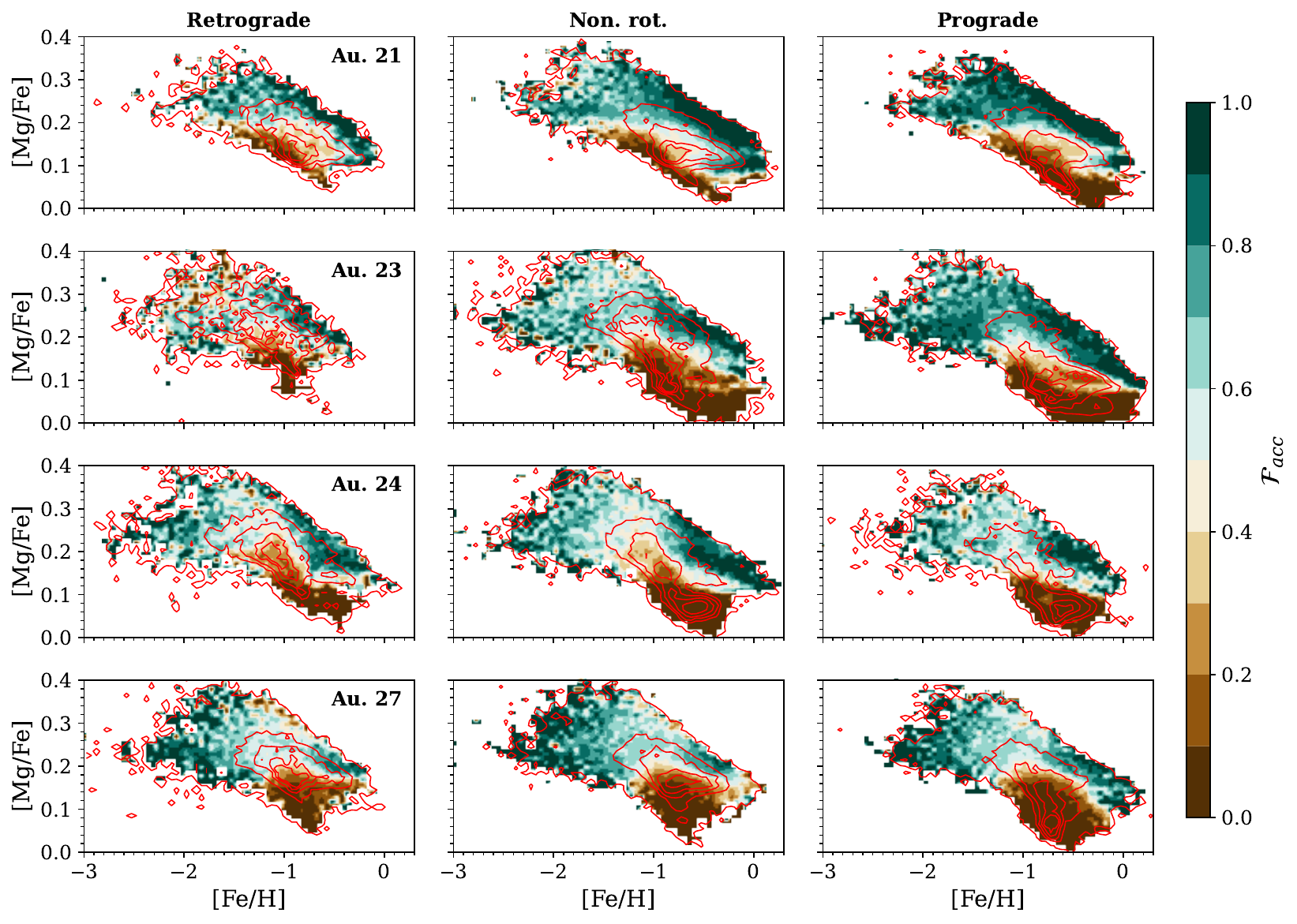}
   \caption{Fraction of accreted stellar particles in different area of the [Fe/H] vs [Mg/Fe] diagram for the retrograde, non-rotating, and prograde particles of the kinematically selected halo in the Solar vicinity of the four simulated galaxies. In each panel, the red contour lines show the distribution of the 1, 25, 50, 75 and 90\% distribution of the particles.}
\label{fig:proret_feh_mg}
\end{figure*}

As mentioned in Sect.~\ref{sec:method}, in \citetalias{lovdal_2022} and \citetalias{dodd_2023}, the number of stars assigned to a cluster is revised by selecting all stars within a Mahalanobis distance of 2.13 from the cluster centre, even if they were not part of the core selection made by the single linkage method, and by excluding all stars beyond this distance. Applying the same method to our simulations, we found that the fraction of stars increased of order $1\%$. The revised selection had a negligible impact on the purity of clusters, however, the recovery rate was lowered, particularly when considering only the particles belonging to a progenitor that dominates a given clump. For example, the recovery fraction for Prog.~3 in Au.~21 dropped from 66\% to 58\%, for Prog.~5 in Au.~24 it dropped from 29\% to 26\%, and for Prog.~4 in Au.~27 it dropped from 73\% to 65\%. Notably, it is in clusters dominated by these progenitors that the purity is the highest. For the other progenitors, the recovery fraction was similar between the two cases. Given the negligible impact of the revised selection on our results and conclusions, we retain the initial selection obtained by the single linkage method for the rest of the paper. 

\subsection{Chemical and age properties of the individual clumps} \label{sec:chemage}

Although, the single linkage method only uses dynamical properties to detect the clusters, chemical and stellar age information can be used in disentangling clusters (or structures) dominated by accreted particles from those mostly populated by stars formed in-situ \citep[e.g.][]{nissen_2010,kruijssen_2020,gallart_2019,belokurov_2022,bellazzini_2023,dodd_2023}. Fig.~\ref{fig:clLZage_feh_mg} shows the median [Fe/H], [Mg/Fe] and stellar age of the clusters' population as function of their average vertical angular momentum, colour-coded by the fraction of accreted particles they containe. We separated the clusters in three main orbital categories: retrograde ($\mathrm{L_z}<-500$~km.s$^{-1}$.kpc$^{-1}$), non-rotating ($-500<\mathrm{L_z}<500$~km.s$^{-1}$.kpc$^{-1}$), and prograde ($\mathrm{L_z}>500$~km.s$^{-1}$.kpc$^{-1}$). In all the simulations, the large majority of clusters that are predominantly composed of in-situ particles $(\mathcal{F}_{acc}<0.3)$ are prograde. Although, as mentioned in the previous section, the detection of such a high number of clusters dominated by in-situ particles is unexpected, it is not surprising that such clusters are prograde. This is consistent with the fact that in the solar vicinity,  50-70\% of the prograde kinematically selected stellar halo is populated by in-situ particles, and that 40 to 60\% of the in-situ particles of the stellar halo are on prograde orbits, except for Au.~24, where this percentage drops to 29\%, as shown in Fig.~\ref{fig:dominance}. 

However, some clusters dominated by in-situ particles are also encountered on non-rotating and on retrograde orbits. The presence of in-situ dominated clusters in the non-rotating region is not unexpected, as in the MW it has be shown that up to $\simeq 50$\% of the halo stars with a low circularity can be born in-situ \citep{bonaca_2017,haywood_2018,dimatteo_2019,amarante_2020}, either because they were formed in the proto-galaxy \citep[also called Aurora;][]{belokurov_2022,chandra_2023}, or because they were part of the disc and were subsequently splashed into halo orbits due to an interaction with a massive progenitor \citep{dimatteo_2019,gallart_2019,belokurov_2020}. What is more surprising is the observation that in-situ particles can also dominate a significant fraction of retrograde clusters ($\geq 60\%$, except for Au.~23), with a contribution of in-situ particles to the cluster in the order of 50-60\%, and in some cases reaching up to $\simeq 90\%$ in Au.~21. This is unexpected, given that the retrograde halo is generally thought to be largely populated by accreted stars \citep{naidu_2020,myeong_2022,horta_2023,ceccarelli_2024}.

From Fig.~\ref{fig:clLZage_feh_mg}, we can see that prograde clusters dominated by in-situ particles tend to be more metal-rich by approximately 0.3~dex compared to those dominated by accreted particles. However, in [Fe/H], some clusters dominated by accreted particles overlap with clusters mostly populated by in-situ particles. When the contribution of in-situ stars to the clusters' properties is excluded, as indicated by the black triangles, the median metallicity of prograde clusters generally tends to be lower and more widely distributed. However, some clusters still exhibit a relatively high metallicity ([Fe/H]~\(\simeq -0.6\)~dex), similar to when in-situ particles are accounted for, even for well-populated clusters ($>100$ particles), particularly in Au.~23 and Au.~27. On the other hand, [Mg/Fe] and stellar age provide a clearer distinction between prograde clusters dominated by accreted particles and those dominated by in-situ particles, with the former ones being $\sim 0.05-0.1$~dex richer in [Mg/Fe] and $\sim 4$~Gyr older than the latter. This distinction is further confirmed by the median values of these quantities when in-situ particles are removed from prograde clusters, as the median [Mg/Fe] and stellar age accounting only for accreted particles closely resemble those of clusters that are actually dominated by accreted particles.

Regarding the non-rotating and retrograde clusters, none of the studied parameters can effectively distinguish between clusters dominated by accreted particles and those dominated by in-situ particles. Indeed, regardless of the galactic origin of the dominant populations, the clusters are populated by old ($\sim$ 10-12~Gyr) metal-poor ($\mathrm{[Fe/H]}<-1.0$) and alpha-rich ($\mathrm{[Mg/Fe]}\simeq 0.2$) particles. Even when in-situ particles are not taken into account, the properties of the non-prograde clusters remain similar. This is in line with the results of  \citet{khoperskov_2023a}, who find that in-situ particles without net rotation and on retrograde orbit have a similar mean metallicity than the accreted particles, while prograde in-situ particles are $\sim 0.5$~dex more metal-rich than accreted one on similar orbits. However, none of the detected clusters dominated by in-situ particles have a median metallicity lower than [Fe/H]$<-1.35$. Although the statistics are limited, this allows us to tentatively suggest that clusters with a median metallicity lower than $\mathrm{[Fe/H]} < -1.35$ are dominated by accreted particles, while the origin of the dominant population for clusters above this threshold remains unclear. Nevertheless, it is important to stress that the majority of clusters with a high fraction of accreted particles have a median metallicity $\mathrm{[Fe/H]} > -1.4$. As such, this criterion does not allow for the detection of all the debris left by accreted galaxies. 

It is not completely possible to exclude the possibility that some of the particles classified as in-situ are actually formed in the gas clouds brought by an accreted galaxy, in particular for those on non-rotating or retrograde orbits \citep[e.g.][]{pillepich_2015} . If it is the case, this might explain why the fraction of in-situ particles on retrograde orbits ($\sim 50\%$) is higher than found in other simulations \citep[e.g.][]{khoperskov_2023a}, but also why the clusters detected on these orbits share the same properties, regardless of the fraction of accreted particles they contain. However, it seems unlikely that this is the only explanation, as the high fraction of in-situ particles found in clusters would require that 40-70\% of the stars originally from an accreted event were formed during or shortly after the accretion process.

It is interesting to compare the chemical characteristics of the clusters' populations with the fraction of accreted particles measured in the [Fe/H]-[Mg/Fe] plane for different orbital properties, as shown in Fig.~\ref{fig:proret_feh_mg}. With the notable exception of Au.~24, the relative fraction of accreted stars remains largely unchanged across different orbital properties. In-situ particles dominate the low-alpha region, while accreted particles dominate the high-alpha region, and [Fe/H] $\lesssim -1.4$, regardless of [Mg/Fe] value. This last point supports our hypothesis that clusters with median metallicity below this threshold are dominated by accreted particles, regardless of its orbital properties. At higher metallicities, a clear separation emerges between in-situ and accreted particles based on [Mg/Fe]. However, the contour lines representing the particle distribution indicate that, that except for the prograde region, the vast majority of the particles have chemical properties where both accreted and in-situ population coexist, with typically an accreted fraction of $\sim 0.3$. This explains why the chemical properties of retrograde and non-rotating clusters are similar, regardless of the origin of their population. 

It is also interesting to note that prograde in-situ particle reach a higher metallicity than that on retrograde and non-orbiting orbit. Accreted particles do not necessarily exhibit such behaviour, as the most metal-rich stars tend to originate from one or two accreted progenitors. These progenitors can have varying orbital properties from one galaxy to another, reflecting the diverse formation histories of these galaxies. This is very interesting as \citet{kordopatis_2020} found super-solar metallicity stars on retrograde orbits in the MW, and the simulations seems to suggest that these stars can have an accreted origin.

\begin{figure*}
\centering
\includegraphics[angle=0,clip,width=18.5cm,viewport=0 0 845 285]{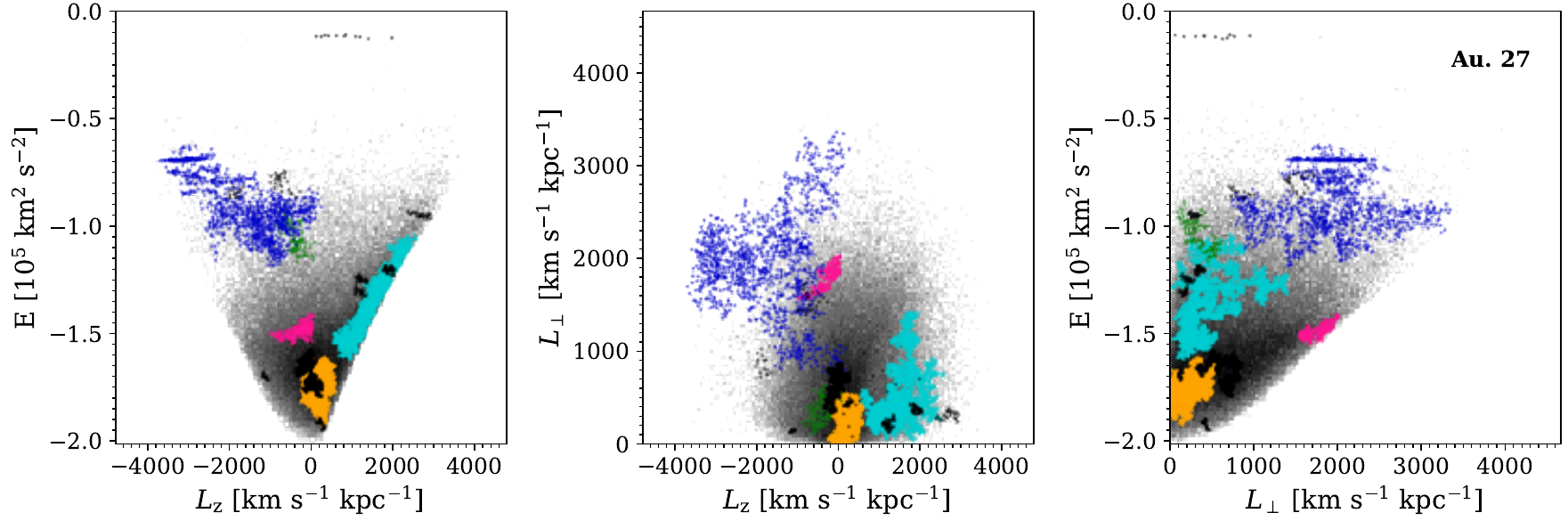}
   \caption{Location in the IOM space of the grouped of clusters found in the kinematically selected halo in the Solar vicinity of Au.~27 using the Mahalanobis distances. The colours scheme used is the same as for Fig.~\ref{fig:dendrogram}. The clusters that are not part of any group are shown in black. The background grey scale show the density variation. } 
\label{fig:IOM_clusters}
\end{figure*}

\begin{figure*}
\centering
  \includegraphics[angle=0,clip,width=18.5cm]{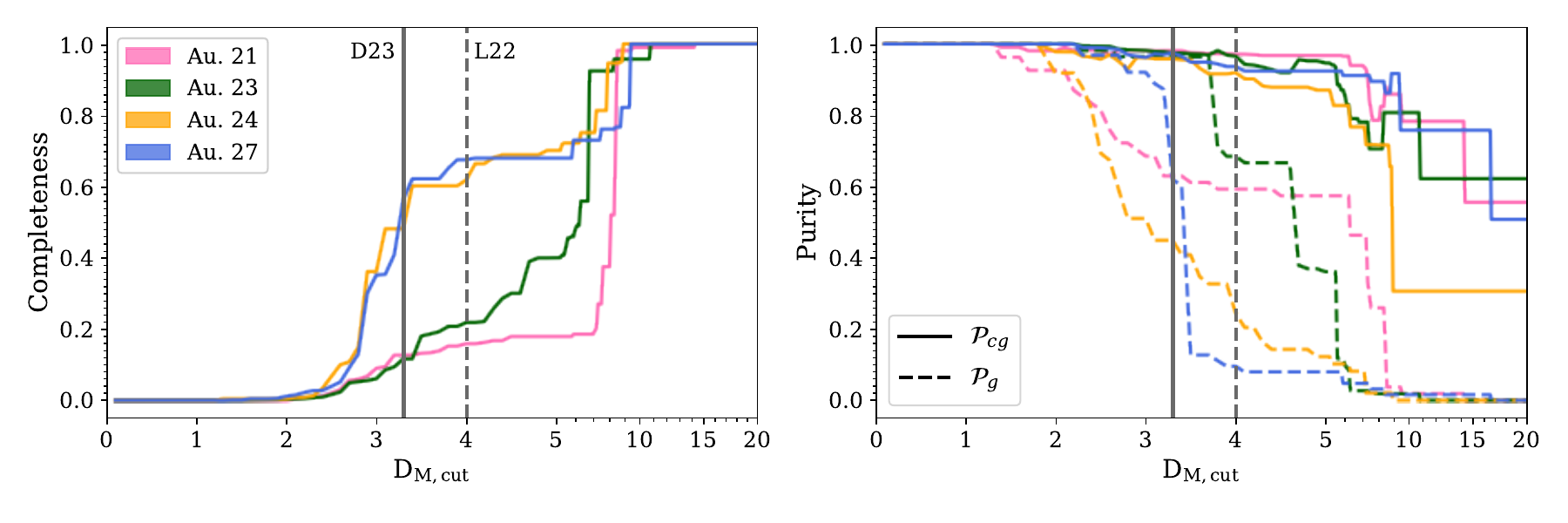}
   \caption{Completeness and purity of the groups of clusters as a function of the  Mahalanobis distance threshold ($\mathrm{D_{M,th}}$) chosen to group clusters together.  For the purity, the solid lines represent the purity defined as the average cluster purity within groups ($\mathcal{P}_{cg}$), while the dashed lines indicate the average purity of the groups ($\mathcal{P}_{g}$). The threshold values used by \citetalias{lovdal_2022} and \citetalias{dodd_2023} are indicated by the vertical dashed and solid lines, respectively. Note that the completeness and purity reported here only refer to the accreted progenitor. Stellar particles formed in-situ are excluded (see Sec.~\ref{sec:GroupClumps}.} 
\label{fig:purcomp_Maldist}
\end{figure*}

\subsection{Association of clumps into groups} \label{sec:GroupClumps}

As discussed in \citetalias{lovdal_2022} and \citetalias{ruiz-lara_2022}, the fact that the single linkage method finds an individual cluster does not necessary mean that it is a unique structure by itself, because a single accreted galaxy can generate several clusters/over-densities in the IoM space \citep{gomez_2013,jean-baptiste_2017,grand_2019,koppelman_2020,khoperskov_2023,mori_2024}. In that sense, \citetalias{lovdal_2022} tentatively proposed that neighbouring clusters in dynamical space might be populated by the same progenitor. As the volume occupied by each cluster is different, these authors used the Mahalanobis distance between two clusters as a metric to find group of nearby clusters:
\begin{equation}
    \mathrm{D_M}= \sqrt{(\bm{\mu_1}-\bm{\mu_2})^T (\Sigma_1+\Sigma_2)^{-1} (\bm{\mu_1}-\bm{\mu_2})},
\end{equation}
\noindent where $\mu_i$ and $\Sigma_i$ describe the mean and the covariance matrix of the $i$-th cluster, respectively. The Mahalanobis distance between two clusters is then used by a single linkage method to generate a dendrogram as shown on the top panel of Fig.~\ref{fig:dendrogram} (for Au.~27;  the same figures for the other haloes are available online). The individual clumps are associated in groups according to a Mahalanobis distance cut-off value. We colour-coded those that are associated to each other adopting the same cut-off as \citetalias{dodd_2023}.
 
In Au.~27, there are five groups of clumps associated with each other, with group sizes ranging from 2 to 25 clusters. This is similar to the number of groups found by \citetalias{lovdal_2022} and \citetalias{dodd_2023} in the MW. Additionally, 10 clusters are not associated with any group because they are too distant from other clusters in the IoM space. This is clearly visible in Fig.~\ref{fig:IOM_clusters}, which shows the locations of these different groups in the IoM space of Au.~27.

The dendrogram of Fig.~\ref{fig:dendrogram} shows that clumps for which the most dominant accreted progenitor is Prog.~1 (in blue) or Prog.~3 (in pink) tend to be grouped together. However, there is no strict one-to-one correspondence between a single group and a single main accreted progenitor. Moreover, we can see that some clusters with different dominant accreted progenitors are grouped together. For instance, cluster~61, which is well-populated ($N_{part}\simeq 380$) and dominated by Prog.~4 (red), is linked to a group that is mostly dominated by particles formed in situ and in Prog.~1 (see Table.~\ref{tab:cl27}).

It is clear from that figure that the cluster associations are highly sensitive to the Mahalanobis distances threshold used to group clusters together. For example, a threshold slightly lower than the one used by \citetalias{dodd_2023} would prevent cluster~61 to be linked to most of the clusters dominated by Prog.~1. On the contrary, using the same threshold as \citetalias{lovdal_2022} will result in more clusters dominated by different accreted progenitors being grouped together. While the specifics vary from halo to halo, this general trend remains consistent across the different simulated galaxies. Therefore, a natural question arises: is there a Mahalanobis distance threshold that can maximize the number of clusters dominated by the same progenitor grouped together without incorporating clusters dominated by other progenitors? In other words, can we optimize both the completeness of clusters dominated by the same progenitor linked into the same group, and the purity of each group of clusters? Given the predominance of in-situ contaminants, and they are not expected to be clustered in IoM space, we do not account for in-situ particles in the following calculations\footnote{The in-situ particles are not taken into account to compute the completeness and the purity but the search for clusters and groups is still performed on both in-situ and accreted particles.}.

We define the completeness as the fraction of pairs of significant clusters dominated by the same progenitor that are grouped together. Mathematically, for a given Mahalanobis distances threshold $(\mathrm{D_{M,th}})$ this is expressed as: 
\begin{equation}
    \mathcal{C}(\mathrm{D_{M,th}})=\sum_p \frac{2}{C_p(C_p-1)} \sum_{\substack{i, j = 1 \\ i \neq j}}^{C_p} \delta(G_i, G_j) \,
\end{equation}
where $p$ represents a given dominant progenitor, $C_p$ is the number of clusters dominated by this progenitor, and $G_i$ is the ID of the group to which belong the cluster $i$. The evolution of the completeness of the four simulations studied here as function of the adopted Mahalanobis distance threshold is presented in the left panel of Fig.\ref{fig:purcomp_Maldist}. We see that for Au.~24 and Au.~27, the completeness rises rapidly, reaching a plateau of 0.7 between $3.5<\mathrm{D_{M,th}}<7.0$. On the contrary, for Au.~21 and Au.23, the completeness is below 0.2 up to $\mathrm{D_{M,th}}\simeq$ 4-5, and then rises suddenly to unity. Several factors can explain the difference observed between the different simulations. For Au.~24, the quick rise of the completeness is explained by the fact that five different progenitors dominate the different clusters, which are furthermore relatively close one to another. Therefore, the distance threshold needed to regroup together most of the clusters dominated by the same progenitor is relatively low. For Au.~23, the slow rise of the completeness is a consequence of the large spread in the IoM of Prog.~1 which is highly clustered at different energy levels (see Fig.~\ref{fig:dominanceAccOnly}). Therefore, a higher distance threshold is needed to link most of them into the same group. The reasons behind the difference between Au.~21 and Au.~27 are less clear. A tentative explanation is that in Au.27 the two progenitors that dominate the IoM (Prog.~1 and 3, see Fig.~\ref{fig:dominanceAccOnly}) occupy two different regions in the IoM space, while at the same time, for Au.~21, the separation between Prog.~1 and 2 is less clear, as they overlap in energy and perpendicular angular momentum. From these results, we observe that with the Mahalanobis distance threshold adopted by \citetalias{dodd_2023}, on average, between 11\% and 56\% of the clusters dominated by the same progenitor are grouped together. With the less restrictive threshold adopted by \citetalias{lovdal_2022}, these values increase to 15\% and 68\%. Based on these results, it is likely that several different groups found by \citetalias{lovdal_2022} and \citetalias{dodd_2023}, which are assumed to be populated by different accreted galaxies, are actually formed by the same single progenitor.

Regarding the purity, several definitions are imaginable. For instance, it is possible to define the purity as the relative contribution of the dominant progenitor of the group in terms of mass fraction (or fraction of particles). In that case, the purity is largely dominated by the most populated cluster of the group, which often is an order of magnitude more populated than the other clusters. As a result,  with this definition, the purity does not change significantly with the Mahalanobis distance threshold adopted. Although this definition of the purity is valid, its use in determining the optimal threshold value is limited. An alternative and potentially more insightful definition of the purity is to consider it as the fraction of clusters within a given group that are dominated by the most common dominant progenitor in that group, averaged over the different groups. In other words, with this definition, the purity corresponds to the average fraction of clusters grouped together that are dominated by the same progenitor. Mathematically, this can be expressed as:

\begin{equation}
\mathcal{P}_{cg}(\mathrm{D_{M,th}})=\frac{1}{\mathrm{G}(\mathrm{D_{M,th}})}\sum_{g=1}^{\mathrm{G}}\left(\frac{1}{C_g} \sum_{i=1}^{C_g}  \delta(P_{gi}, P_{g,most}) \right)
\label{eq:purity1}
\end{equation}
where $\mathrm{G}(\mathrm{D_{M,th}})$ is the number of groups for a given Mahalanobis distance threshold value $(\mathrm{D_{M,th}})$, $C_g$ is the number of clusters belonging to the group $g$, and $P_{gi}$ is the dominant progenitor of the cluster $i$ of the group $g$. The term $P_{g,most}$ represents the most common progenitor in a group $g$, defined as:

\begin{equation}
    P_{g,most}=\mathrm{argmax}_P \left( \sum^{C_g}_{i=1} \delta (P_{gi},P) \right) \ .
\label{eq:Pmost}
\end{equation}
Note that in the eventuality of several equally dominant progenitors, we randomly select one of the progenitors, as the result of Eq.~\ref{eq:purity1} is invariant to the choice of the dominant cluster in Eq.~\ref{eq:Pmost}. With this definition, when $\lim{ \mathrm{D_{M,cut}} \to\infty}$, the purity reaches a plateau that corresponds to the fraction of clusters that are dominated by the same progenitor (i.e. the dominant progenitor the most encounter across all the clusters). As visible from the plain lines in the right panel of Fig.~\ref{fig:purcomp_Maldist}, we observe that the most frequent dominant progenitor tends to dominate 60\% of the overall clusters found in the halo of the Solar vicinity, except for Au.~24, where it dominates only 30\% of the clusters. The reason for this is that in the other galaxies, two progenitors dominate 80\% of the clusters, while in Au.~24, the number of dominant progenitors is higher. It is also noticeable that, with this definition, the evolution of the purity is not monotonic, and can increase when $\mathrm{D_{M,th}}$ increases. Although this might seem counter-intuitive, it highlights the fact that two groups of different sizes dominated by the same progenitor can be joined together, which in some cases can increase the purity. Using this definition, we see that the purity is relatively similar for both Mahalanobis distance threshold adopted by \citetalias{lovdal_2022} ($\mathcal{P}_{cg}\simeq 0.94$) and by \citetalias{dodd_2023} ($\mathcal{P}_{cg}\simeq 0.97$). This mean that, respectively, on average, only 6\% and 3\% of the clusters of a group have a different dominating progenitor than the other clusters of the group. However, this number has to be taken with care, as at low $\mathrm{D_{M,th}}$, the large majority of the group are composed of 2 or 3 clusters, which can bias the results. Instead, we find that the purity is above 0.9 for groups composed of maximum four clusters, and decrease to 0.6-0.7 for group of six or more clusters.

\begin{figure*}
\centering
\includegraphics[width=0.7\textwidth]{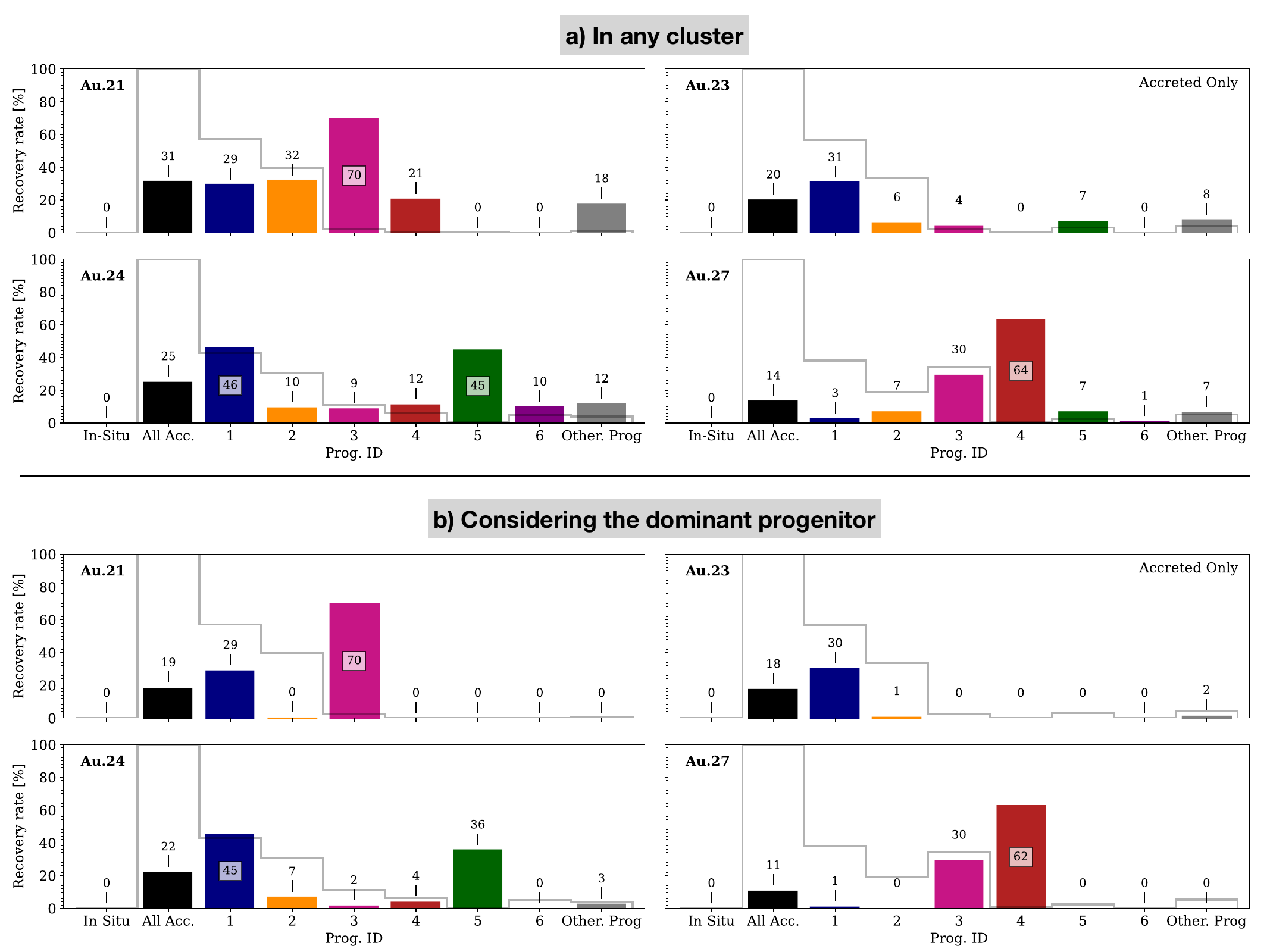}
   \caption{Same as Fig.~\ref{fig:recov_ISacc} but considering only the accreted particles. } 
\label{fig:recov_acconly}
\end{figure*}

\begin{figure*}
\centering
  \includegraphics[angle=0,clip,viewport=0 130 1025 754,width=17cm]{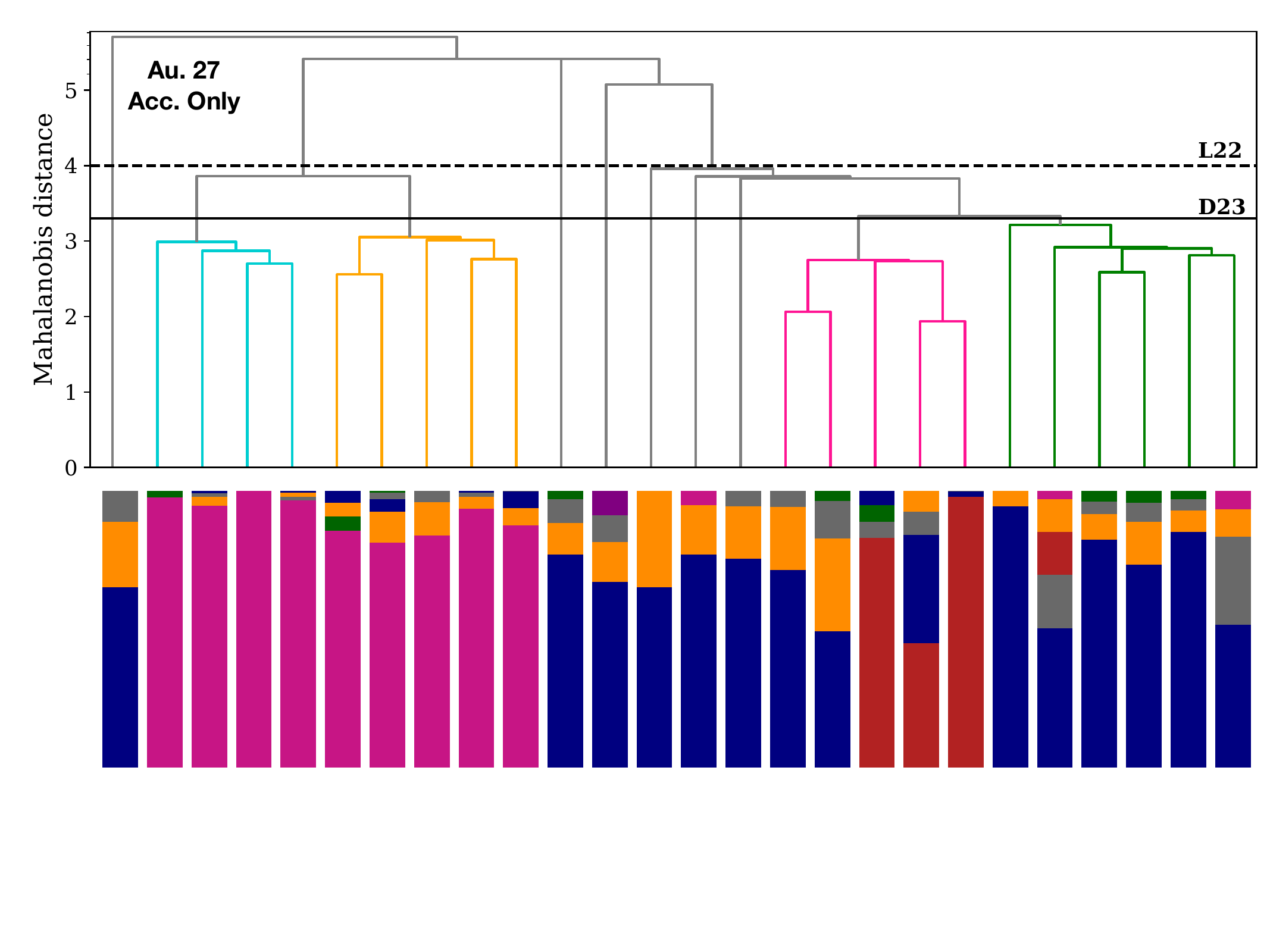}
   \caption{Same as Fig.~\ref{fig:dendrogram} but for the Solar vicinity sample, considering only the accreted particles for Au.~27. } 
\label{fig:dendrogramAcc27}
\end{figure*}

Another interesting way to define the purity is to wonder what is the probability that a group is entirely composed of clusters dominated by the same progenitor. It is mathematically defined as:
\begin{equation}
\mathcal{P}_g(\mathrm{D_{M,cut}})=\frac{1}{\mathrm{N_c}} \sum_{g=1}^{\mathrm{G(\mathrm{D_{M,cut}})}} \sum_{i=1}^{\mathrm{C_g}} \delta(P_{g1},...,P_{gi}) \ ,
\label{eq:purity2}
\end{equation}
where $\mathrm{G}(\mathrm{D_{M,cut}})$ is the number of groups for a given Mahalanobis distances cut-off value $(\mathrm{D_{M,cut}})$, $C_g$ is the number of clusters belonging to the group $g$, $P_{gi}$ is the dominant progenitor of the cluster $i$ of group $g$. The evolution of the purity using that definition as a function of the adopted Mahalanobis distance threshold is shown by the dashed lines in Fig.~\ref{fig:purcomp_Maldist}. With this definition, the purity is a monotonically decreasing function that varies from $\mathcal{P}_g\left(\lim{ \mathrm{D_{M,cut}} \to 0}\right)=1$ to $\mathcal{P}_g\left(\lim{ \mathrm{D_{M,cut}} \to \infty}\right)=0$. In that case, the evolution of the purity vary strongly from galaxy to galaxy, with a relatively similar bimodality, likely caused by the same reasons, as seen as with the completeness, with Au.~21 and Au.~23 having a different trend than Au.~24 and Au. 27. Adopting a threshold similar to \citetalias{dodd_2023}, we see that typically between 40 and 60\% of the groups are entirely composed of clusters dominated by the same progenitor, while with the threshold used by \citetalias{lovdal_2022} this fraction can decrease significantly: to $\sim 15$\% for haloes Au.24 and Au.~27.

In conclusion, it is challenging to determine an optimal Mahalanobis distance threshold that can simultaneously optimise both the purity and completeness of the groups of clusters. This difficulty arises from the significant variation in the results obtained from galaxy to galaxy, which reflects the diversity of their accretion histories and of the properties of their accreted galaxies. Nevertheless, following the assumption we already presented that Au.~27 is the simulations with the closest properties to the MW, it seems that a threshold close to the value adopted by \citetalias{dodd_2023} of $\mathrm{D_{M,cut}}=3.3$ is well suited. Indeed, in that case, the average purity of the clusters in groups ($\mathcal{P}_{cg}=0.97$) and the purity of the groups themselves ($\mathcal{P}_g=0.62$) is relatively high, with a completeness of 57\%\footnote{We remind here that the purity and completeness are calculated only based on the accreted particle. However, the clusters and the groups are identified using both in-situ and accreted particles.}. A higher value, such as that adopted by \citetalias{lovdal_2022} will lead to a slightly higher completeness, but decrease the purity of the group by more than a factor of 6. \citetalias{ruiz-lara_2022} suggested that the metallicity distribution function (MDF) and the colour distribution could be used to increase the purity of the groups by identifying outliers. This approach could also improve the completeness by linking clusters with different dynamical properties but similar metallicity distributions, as could be the case for their Cl.~62 that has a similar MDF to Thamnos 1 \& 2, despite its locus in a different region of IoM space. We applied a similar method to the MDF and the age distribution function (ADF), but did not observe improvements in either the purity nor the completeness. This lack of improvement may be attributed to the high contribution of in-situ particles within the detected clusters, and potentially to the misclassification of in-situ particles that originated from accreted gas clouds. Consequently, we chose not to explore this method further at this stage and will defer its analysis to the following section, where we focus on a sample composed solely of accreted particles.

\begin{figure*}
\centering
  \includegraphics[angle=0,clip,,width=17cm]{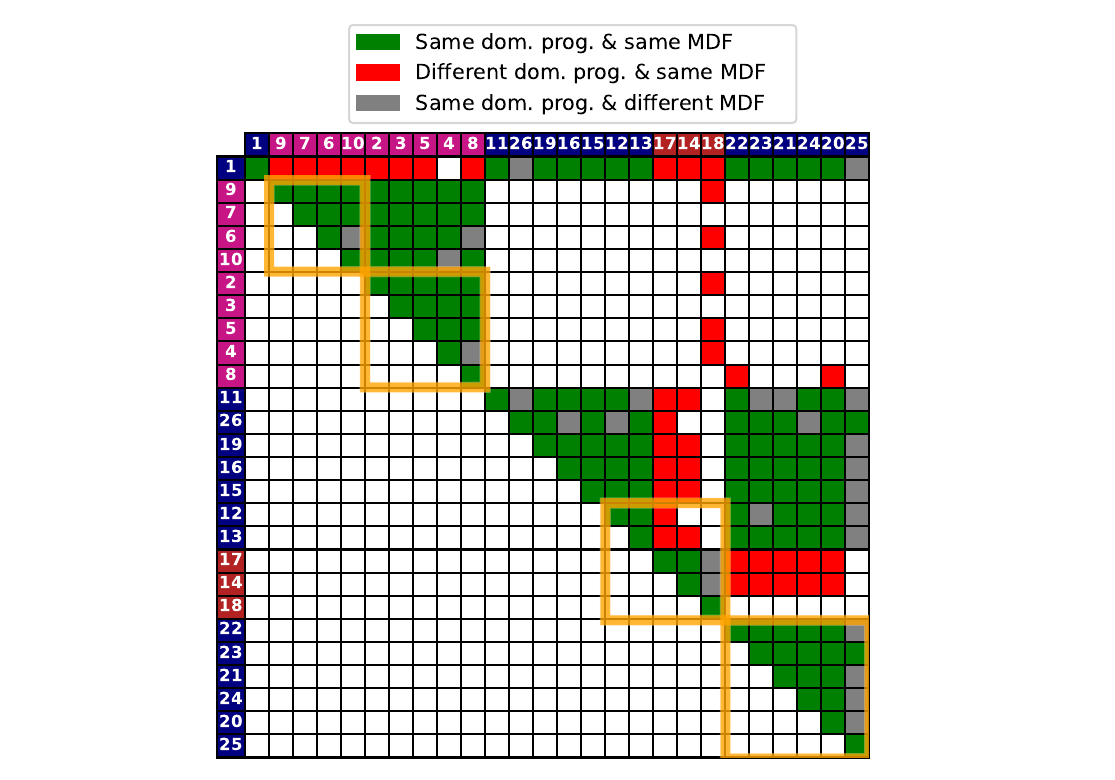}
   \caption{Confusion matrix of the Kolmogorov-Smirnov test, comparing the MDF of each significant cluster found in Au.~27 considering only particles from accreted progenitors. The green cells indicate pairs of clusters sharing the same dominant progenitor and which have a similar MDF at the 95\% confidence level (p-value >0.05, true positive). The red cells show pairs of clusters with a similar MDF but with different dominant progenitor (false positive). The grey cells show the pair of clusters with the same dominant progenitor but with different MDF at the 95\% confidence level (false negative). The pair of clusters that are not dominated by the same progenitor and that have not the same MDF at the 95\% confidence level (true negative) are shown by the white cells. As the matrix is diagonally symmetric, we decided to not colour code the cells in the lower part of the matrix for visibility reasons. The orange rectangles show the groups of clusters linked together by the Mahalanobis distance using the same cut-off value of \citetalias{dodd_2023}. The clusters are ordered in the same way that in Fig.~\ref{fig:dendrogram}, with their reference ID. indicated in each row and column.} 
\label{fig:KSMetacc}
\end{figure*}

\begin{figure*}
\centering
  \includegraphics[angle=0,clip,width=17cm]{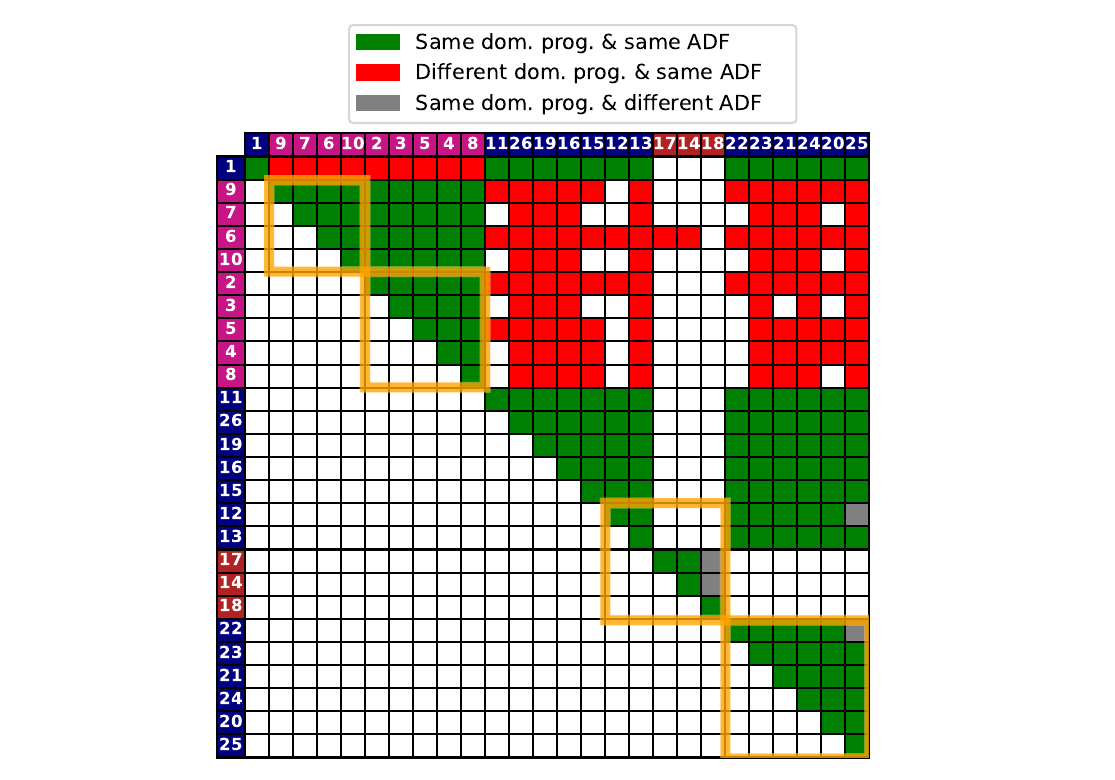}
   \caption{Same as Fig.~\ref{fig:KSMetacc} but using the age distribution function (ADF) instead of the MDF.} 
\label{fig:KSAgeacc}
\end{figure*}

\subsection{Analysis of a pure sample of accreted particles} \label{sec:acconly}

In the above analysis, we have seen that the in-situ component often dominates the budget of particles that belong to statistically significant clusters. In this section, we apply our methods to a accreted particles only\footnote{Here we drop the cut in velocity to select particles with halo-like kinematics.}. This scenario is, of course, optimistic, as even the most advanced methods proposed in the literature for separating in-situ and accreted stars in observed samples do not achieve 100\% purity and completeness. 
For example, chemistry such as [Al/Fe] and [Mg/Mn] may not be sufficient to disentangle accreted and in-situ stars with [Fe/H]$< -1.3$, despite the relatively successful separation for more metal rich stars \citep{hawkins_2015,belokurov_2022,das_2020,horta_2021,fernandes_2023}.

\subsubsection{Properties of the significant clusters}

With the notable exception of Au.~24, the number of clusters found without the inclusion of the in-situ particles is usually between 50 and 60\% lower in the same IOM region than previously. Interestingly, across the different galaxies, this ratio does not change significantly between the retrograde, non-rotating and prograde regions. This confirms the important role played by the in-situ particles in the clusters' detection, as we already see in Section.~\ref{sec:recovrate}. 

Interestingly, we note that even when the in-situ particles are removed, none of the clusters are perfectly pure, as there is always some overlap between different progenitors. However, the contamination is relatively low, as on average the dominant progenitor contributes to 70 to 80\% of the particles of a cluster. However, as with the inclusion of in-situ particles, we do not observe a strong dependence between the purity and significance of the clusters, nor with the number of particles that belong to a cluster.

Figure~\ref{fig:recov_acconly} shows that the recovery rate of individual progenitors is in general higher after excluding in-situ stars. In particular, when considering only the particles belonging to clusters where their progenitor is the dominant contributor (bottom panel), some progenitors whose recovery fraction was very low in the in-situ+accreted case now become much more prominent. This is particularly the case for the progenitor that contribute the most to the stellar halo in the Solar vicinity and/or to the prograde region (for example, Prog.~1 in Au.~21, Prog.~1 in Au.~24 and Prog.~3 in Au.~27). In addition, the percentage of particles proceeding from smaller progenitors that are picked up as part of clusters also increases. This is not surprising as in-situ stars are smearing some of the clumpy features out. 

Even in the best case scenario of dealing with accreted particles only, for most of the progenitors, only a minority of particles are found to reside in statistically significant clusters. As for the case when in-situ particles are included, the progenitors with the higher recovery rates are those that have been most recently accreted ($<6$~Gyr ago), as they are the most clumpy in IoM space (see Fig.~\ref{fig:dominanceAccOnly}). This essentially implies that a complete census of stars belonging to a given progenitor is hard to obtain with this methodology, even in the very favourable case of an unbiased target selection, only limited in volume. Furthermore, our simulations do not include uncertainties in distances, proper motions and radial velocities, whose effect would clearly blur the picture even more by decreasing the significance of the clusters and increasing the contamination fraction \citep[see e.g.][]{helmi_2000}.

\subsubsection{Properties of the groups of clusters}

What about grouping clusters together? Fig.~\ref{fig:dendrogramAcc27} exemplifies the results for Au.~27. While in this case the majority of clumps have a clear dominant progenitor, and some clusters are well grouped together in the first passes of the single-linkage through the Mahalanobis distance, it is difficult to find a common threshold that gives the desired result for each group. For example, a threshold of $\mathrm{D_{M,th}}=4$ would group together all the clusters where Prog.~3 is dominant and the majority of those containing Prog.~1; at the same time, clusters related to Prog.~4 would be ingested in the group where Prog.~1 is dominant in most clusters. A lower threshold of $\mathrm{D_{M,th}}=3.5$ would instead produce an overestimated number of groups by forming two separate groups dominated by Prog.~3, and an even lower threshold of $\mathrm{D_{M,th}}=3$ would not link the majority of the clusters dominated by Progs.~1 and 4 to any group. While each simulated galaxy has its own specificity, these general trends are common across the haloes analysed. For comparison with the observations, we find that excluding in-situ particles and using the \citetalias{dodd_2023} threshold, the completeness is $\sim 0.25$ in all the simulations, compared to $\sim 0.6$ in Au.~24 and Au.~27 when in-situ particles are included. On the other-hand, the purity does not change significantly, regardless of the definition used, except for Au.~24 where the purity of the groups goes up to $\mathcal{P}_g=0.64$ (compared to 0.45 previously). To reach a completeness of 0.57 in Au.~27, i.e. similar to the value found with the inclusion of in-situ particles in Sect.~\ref{sec:GroupClumps} using the threshold of \citetalias{dodd_2023}, the Mahalanobis distance threshold have to be increased to 3.9. In that case, the average cluster purity within groups is of $\mathcal{P}_{cg}=0.96$ and the average purity of the groups is of $\mathcal{P}_g=0.50$, slightly lower than when the in-situ stars are included. 

As mentioned earlier, \citetalias{ruiz-lara_2022} used the MDF and the colour distribution of clusters to help increase purity and completeness of their progenitors. Several other studies have also used, either partially or exclusively, chemical properties to identify groups of stars \citep[e.g.][Bokyoung in prep.]{horta_2021,horta_2024,naidu_2020,naidu_2021}, or of globular clusters that originate from the same accreted galaxy \citep[e.g.][]{kruijssen_2019,massari_2019,massari_2023}. Since we are working with stellar particles, we have access to the stellar age of individual particles, and we therefore decided to use the age distribution function (ADF) instead of the colour. Following the procedure of \citetalias{ruiz-lara_2022}, we compare the MDF and the ADF of each cluster with the others using a Kolmogorov-Smirnov (KS) statistical test. We assume that the distributions between two clusters are compatible when the KS p-value $P(\phi_1,\phi_2) > 0.05$ for both the MDF and the ADF.

As shown in Fig.~\ref{fig:KSMetacc}, the comparison between the MDF of the clusters reveals that approximately 60\%\footnote{The values mentioned here refer to the fraction of true (false) positives (negatives) compared to the total number of positive (negative) links, as determined by the p-value.} of the cluster pairs with p-values $> 0.05$ are true positives (i.e., clusters dominated by the same progenitor; green cells). In the case of Au.~23, this rate increases to 95\% due to most clusters being dominated by Prog.~1. For the other haloes, about 75\% of the negative MDF comparisons are true negatives (i.e., pairs of clusters not dominated by the same progenitor and with p-values $< 0.05$; white cells). In about 25\% of cases, this method yields false negatives (i.e., exclusion of clusters with the same progenitor; grey cells), and in roughly 40\% of cases, it produces false positives (i.e., associating clusters not dominated by the same progenitor but with similar MDFs; red cells). Interestingly, we systematically detect false positives within groups across all simulations (highlighted by the orange rectangles). In some cases, two clusters with a similar MDF show a significant contribution from the dominant progenitor of the other cluster in the pair (e.g., Cls.~59, 60, and 74 of Au.~23). However, in other cases, clusters grouped together with a similar MDF are not populated at all by stars from the same progenitors (e.g., in Au.~27 with Cl.~17, which is dominated by Prog.~4, and where Prog.~1 contributes only 5\% of the cluster population, and Cls.~12 and 13, which are dominated by Prog.~1 and contain no particles from Prog.~4). Furthermore, this phenomenon cannot be solely attributed to small number statistics, as false negatives occur also in clusters populated by several hundred particles (e.g., Cls.~4 and 18 of Au.~27, although in that case, the clusters are not dynamically grouped).

The comparison of the ADF presented in Fig.~\ref{fig:KSAgeacc} shows a similar ratio of true positives, generally around $\sim 60\%$, except for Au.~23. However, the ADF comparison appears to be less prone to false negatives (grey cells), with a true negative ratio of approximately 90\%. When comparing the ADF of individual clusters, a mix of true negatives and false positives emerges, regardless if the clusters are associated or not into dynamical groups. Across all simulations, this behaviour seems to complement the MDF results, though this is not systematic (e.g., Cl.~23 in Au.~24). Thus, by using both the MDF and ADF, it may be possible to identify intruder clusters that are not dominated by the same progenitor as the other clusters in the group, as well as to find groups of clusters potentially populated by the same progenitor. This suggests that a promising approach could be to directly compare the age-metallicity relation, as was recently done by \citet{dodd_2024} with Sequoia and Thamnos.

\begin{figure}
\centering
  \includegraphics[angle=0,clip,width=8.5cm]{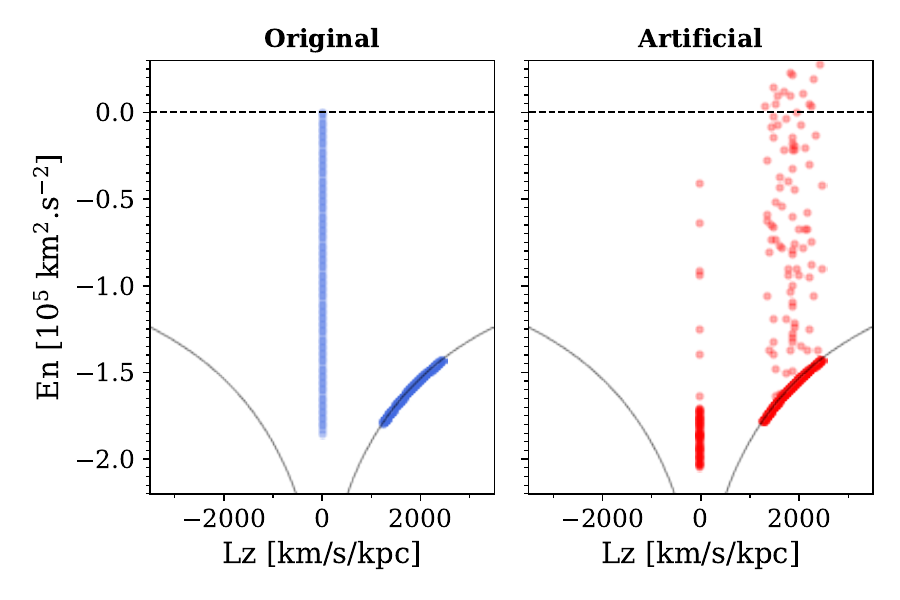}
   \caption{Toy model illustrating the over/underdense regions in the IoM of an artificial halo (right panel) created by reshuffling the velocities of the original particles (left panel). In the original distribution, 1000 particles are on perfectly circular orbits and 100 particles on perfectly vertical orbits. In both panels, black lines delineate the possible distribution limits in energy-angular momentum, and the dashed line indicates the energy limit for a particle to be considered bound.} 
\label{fig:toymodel}
\end{figure}

\begin{figure*}
\centering
  \includegraphics[angle=0,clip,width=18cm]{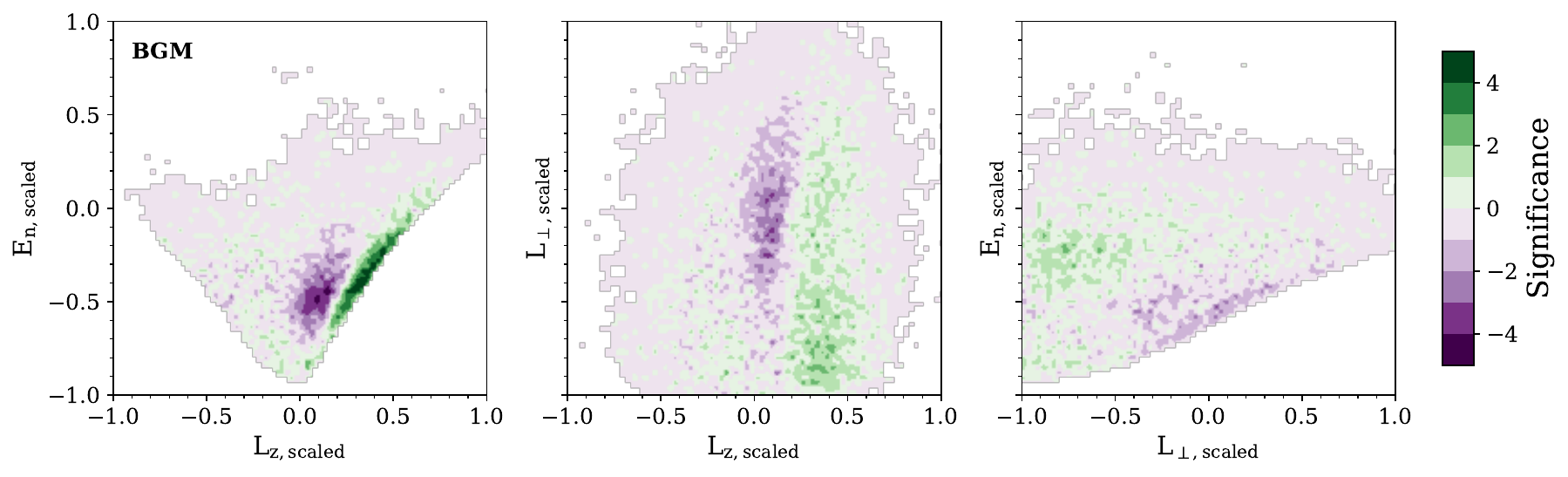}
  \includegraphics[angle=0,clip,width=18cm]{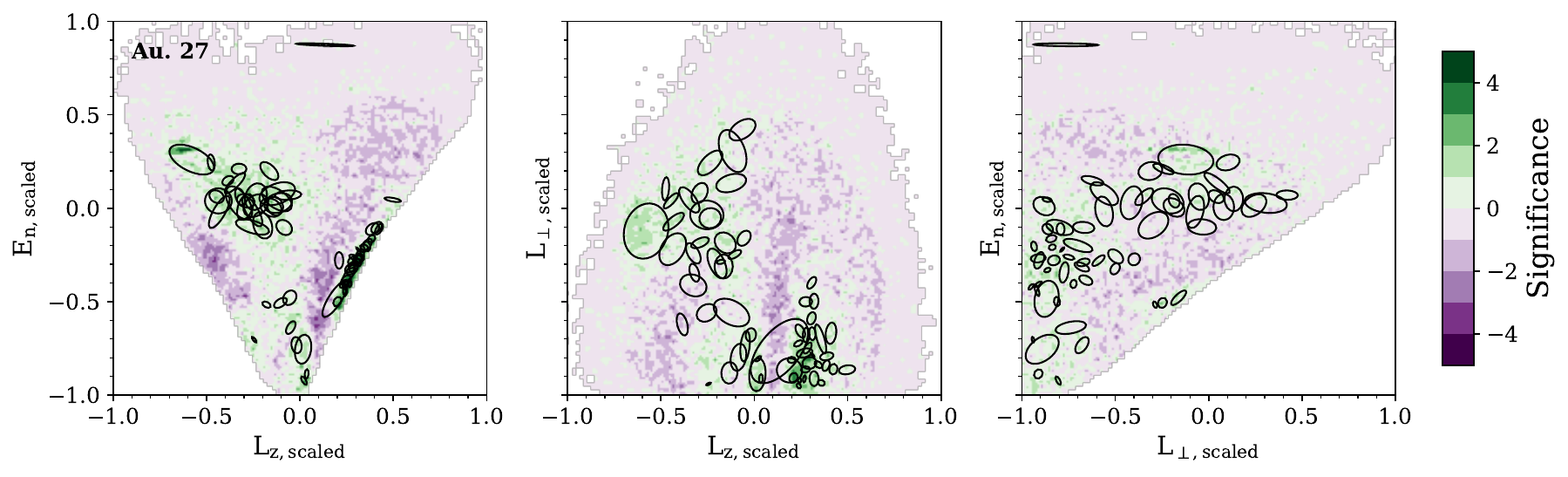}
   \caption{Variation of the significance in the IoM of the originally kinematically selected halo in the Solar vicinity of the Besan\c{c}on Galactic model (upper panels) and of Au.~27 (lower panels) compared to the set of 100 artificial haloes. In the lower row, the ellipses indicate the location of the significant clusters detected by the single linkage method using the in-situ and accreted particles.} 
\label{fig:BGM}
\end{figure*}

\section{The impact of the artificial background haloes} \label{sec:backgen}

Several hints in the previous sections suggest that the cluster detection method might generate artificial significant clusters that do not accurately reflect the actual structure of the local stellar halo. First, unlike the debris of accreted galaxies, in-situ particles are not expected to be particularly clustered in the IoM. It is true that resonances caused by non-axisymmetric features such as the bar or spiral arms can form overdensities in the IoM \citep{dillamore_2024}. However, these overdensities are expected to be relatively large, not well clustered—particularly in $L_\perp$—and mostly found in the prograde region. Despite this, we detected significant clusters mostly populated by in-situ particles in all the simulations (see Sect.~\ref{sec:recovrate}), not only in those where the main galaxy is barred \citep[see Table.~2 of][]{grand_2024}. Furthermore, we detected clusters dominated by in-situ stars not only in the prograde regions but also on non-rotating and on retrograde orbits (see Sect.~\ref{sec:chemage}).

Secondly, we do not observe a strong correlation between the significance and purity of the clusters, as one might naturally expect if the significant clusters were primarily composed of particles from a single accretion event. Additionally, this lack of correlation cannot be only attributed to the presence of misclassified in-situ particles born in accreted gas clouds since even when in-situ particles are excluded, this correlation between the purity and significance of the clusters is not observed.

Finally, if a significant cluster is predominantly populated by particles from a single accretion event, one would expect that the purity should increase if the selection is restricted to the inner part of the cluster, as the contamination should gradually dominate (in terms of relative fraction) further away from the centre of the cluster. However, we did not observe a significant change in the purity of the clusters when restricting the selection to a maximum Mahalanobis distance of 2.13 from the cluster centre, compared to the original and less restrictive selection (last paragraph of Sect.~\ref{sec:recovrate}).

Here, we explore whether some of the detected clusters might be artificially originated by the method used to create the background haloes. As discussed in Sect.~\ref{sec:artHalo}, these background haloes are generated by reshuffling the velocities of the real halo. However, this method does not adequately preserve the large-scale structure of the IoM space, and can instead lead to the creation of local over/underdensities in the IoM space of the artificial haloes compared to the original ones. This issue is clearly visible in the toy model presented in Fig.~\ref{fig:toymodel}. The original halo in this simple model consists of 1,000 particles, equally spaced along the 3D Cartesian X-axis between a radius of 5.5 and 10.5~kpc (i.e., $\pm 2.5$~kpc from the Sun located at $R_\odot= 8$~kpc). These particles are on perfectly circular orbits within an axisymmetric Galactic potential well described by the \citet{mcmillan_2017} potential. Additionally, 100 particles, all located at $X=8$~kpc and equally spaced in energy, are modelled with velocity only along the vertical axis\footnote{The conclusions drawn are similar if these particles are on perfectly radial orbits, except that the angular momentum in the artificial halo tends to be less spread.}. As shown in this figure, the distribution in energy-angular momentum of the artificial halo is locally very different compared to the original distribution. The velocity reshuffling process creates noticeable discrepancies in the IoM  distribution compared to the original distribution. In the artificial halo, we clearly observe the appearance of underpopulated regions, particularly in the non-rotating, high-energy areas, and overpopulated regions, especially in the prograde high-energy region. While the specific location of these over/underdensities depends on the specific IoM distribution of the original halo, this extremely simplistic model clearly highlights the limitations of the reshuffling method in creating artificial haloes and suggests that it may contribute to the detection of artificial clusters that do not accurately reflect the true structure of the local stellar halo.

To investigate further where one can expect this artificial structures to be located, we decided to use a more realistic, yet smooth model of the MW, using the Besan\c{c}on Galactic Model \citep[BGM][]{robin_2012,fernandez-trincado_2017}. We selected stellar particles from the Solar vicinity using the same criteria as \citetalias{lovdal_2022} and \citetalias{dodd_2023} to select the kinematic stellar halo and to generate the set of 100 artificial background. We computed the local significance of the original halo in each pair of IoM parameters by comparing the number of stars in the original halo to a set of 100 artificial haloes, as explained in Sect.~\ref{sec:method}. The creation of the over/underdense regions in the artificial haloes corresponds to the regions with negative/positive significance in Fig.~\ref{fig:BGM}, respectively. When compared to Au.~27, we see a clear similarity in the location of regions of high and low significance between the two models. In particular, in the prograde region (L$_z>0$), we observe that at high vertical angular momentum, the significance is very high, indicating underpopulated artificial haloes, and a region of lower angular momentum where the significance is negative, indicating overpopulated artificial haloes. Interestingly, in the retrograde halo (L$_z<0$), we observe the inverse trend, with a region of positive significance at low energy and low vertical angular momentum, although the absolute amplitude of the significance is lower, likely due to the smaller number of particles in this region. We can observe some similarities between the regions of high significance in these two models and the locations of some significant clusters detected by \citetalias{lovdal_2022} and \citetalias{dodd_2023} in the MW, and conversely, between the low-significance regions and the absence of significant clusters. These analyses raise the question of the reality of some detected structures in the observations.

It would be interesting to explore alternative methods for generating artificial backgrounds that avoid the biases of the current approach. Some potential directions include modelling the background halo based on an assumed potential and distribution function that reproducs well the MW's global properties (e.g. BGM or a self-consistent model based on \citealt{mcmillan_2017}), or using a multi-Gaussian model to describe the IOM space of the background. However, developing and testing such methods is beyond the scope of this paper, which focuses on quantifying the performance of the current technique. Nevertheless, the work presented here provides a useful metric for evaluating the performance of these alternative approaches.

\section{Discussion and conclusions} \label{sec:conclusion}

In this paper, we have analysed the performance of one of the most advanced methods for identifying the debris of past accreted galaxies in the local stellar halo by applying it to four Milky Way-like galaxies from the Auriga simulation suite. This method, described in detail in \citetalias{lovdal_2022}, aims to identify significant clumps of stars in the IoM space by comparing the local number of stars within a predetermined region with a suite of $N$ artificial background haloes.

Although this method is very promising and has proven its capability to identify both known and new structures \citep{lovdal_2022,ruiz-lara_2022,dodd_2023}, we demonstrate in Sect.~\ref{sec:backgen} that the method used to generate the artificial background haloes can produce local density variations that are not representative of the smooth background distribution in the original halo. This can ultimately lead to an artificial boost of the significance for some clusters, or even to the detection of artificial significant clusters, particularly in highly prograde or retrograde-low energy regions, while potentially preventing the detection of significant clusters in other regions.

Setting aside this potential source of error, we found that the method of detecting significant clusters is also limited and does not allow for the recovery of most of the debris left by accreted galaxies. Indeed, in Sect.~\ref{sec:recovrate}, we observed that this method is particularly effective in identifying most of the stellar particles originating from a progenitor that was recently accreted (within the last 6–7 Gyr). However, it fails at recovering most of the debris left by the older accretions, regardless of the actual mass of the progenitor. In fact, most of the particles from these older accretions detected as being part of any cluster tend to actually contaminate the debris left by other accreted galaxies or by particles formed in situ. Even when the detection of the clusters is done only with accreted particles, the method still largely identifies the debris left by the most recent accretions, but debris left by older accretions remain mostly undetected (see Sect.~\ref{sec:acconly}). The only exception is that, in this case, it also identifies some particles originating from the accreted galaxy that contributes the most to the local stellar halo. Nevertheless, other accreted galaxies are mostly not detected. This suggests that in a cosmologically motivated context, debris from the oldest accretion does not stay strongly clumped in the IoM in the Solar vicinity. As mentioned in Sect.~\ref{sec:method} the debris left by accreted galaxies with stellar mass of $\sim 10^6$~M$_\odot$ might not be detected by the algorithm  due to the imposed minimum number of cluster members and the simulations' resolution. However, the observed bias favouring the detection of more recent accretion events should still apply to these smaller progenitors, since low-mass accreted galaxies passing through the Solar vicinity are more susceptible to disruption than their more massive counterparts.

We also found that the vast majority of the significant clusters in the IoM space identified by this method are highly contaminated by in-situ particles, when these last ones are not the most important population. Additionally, almost all the clusters are populated by stars from more than one accreted galaxy, with clusters where $90\%$ of the stars originate from the same progenitor being very rare. Even with the inclusion of chemical and stellar age information (see Sect.~\ref{sec:chemage}), we found it challenging to disentangle clusters dominated by in-situ particles from those dominated by accreted particles. The distinction is clearer for prograde clusters, where in-situ clusters are more metal-rich, younger, and more [Mg/Fe] depleted than their accreted counterparts. However, for clusters with $L_z<500~\mathrm{km~s^{-1}~kpc^{-1}}$, the distinction becomes less clear, as both in-situ and accreted particle-dominated clusters exhibit similar behaviour in metallicity, [Mg/Fe], and stellar age. We tentatively suggest that all clusters with a median metallicity lower than [Fe/H] $<-1.35$ are dominated by accreted particles. Nevertheless, in the few clusters identified below that limit, we found that the contamination from in-situ particles is still as high as $30\%$, although this conclusion is drawn from a very limited sample of clusters. Furthermore, most of the clusters dominated by accreted particles have a metallicity higher than this, although this is also partially explained by the fact that smaller accreted galaxies are not well resolved by the Auriga simulations, and we can expect that in reality, the ratio of cluster with [Fe/H]$<-1.35$ might be more important.

It is possible to group together clusters closed to each other in the IoM space, as done by \citetalias{ruiz-lara_2022} and \citetalias{dodd_2023} in the observations. However, as presented in Sect.~\ref{sec:GroupClumps}, the purity of these groups decreases rapidly, by linking together clusters dominated by different progenitors. The simulations suggest about a third of the groups found by \citetalias{dodd_2023} might consist of clusters dominated by different accreted galaxies. Furthermore, even in such a small region as in the Solar vicinity, the debris of a single old accreted galaxy are spread over a wide region of the IoM space. As a result, several distinct groups can be mostly populated by particles originating from a single accreted galaxy. Our results suggest that up to 40\% of the clusters not grouped together using the threshold of \citetalias{dodd_2023} to identify groups might actually be coming from the same accreted galaxy, although we see a great variation of this value from galaxy to galaxy. 

\citetalias{ruiz-lara_2022} proposed that clusters from the same galaxy can be identified and distinguished from those originating from different galaxies by comparing their metallicity distributions. In Sect.~\ref{sec:acconly} we show that metallicity alone is insufficient for this purpose, as clusters dominated by different galaxies can have similar distribution of metallicity, while clusters dominated by the same galaxy can have different distribution of metallicities, even within the same dynamical group. This last point aligns with expectations of metallicity gradients with energy in accreted galaxies \citep{amarante_2022,horta_2023,mori_2024}. Our analysis clearly shows that such a gradient is to be expected, even in such a small volume around the Sun. A similar pattern is observed when comparing age distributions. However, our results suggest that combining age and metallicity distributions may effectively disentangle clusters dominated by similar/different progenitors. This is particularly promising given the recent availability to derive precise age-metallicity relations for populations located within the Solar neighbourhood \citep[e.g.][Fernandez-Alvar, in.prep]{gallart_2024}, even for a limited sample of stars \citep{dodd_2024}. 

Applying these conclusions to structures identified in the local stellar halo of the MW allows us to hypothesise about their independence, reality, and progenitor characteristics, and to revisit the MW accretion history.

A first hypothesis can be made concerning the group formed by Sequoia \citep{myeong_2019} and L-RL64/Antaeus \citep{lovdal_2022,ruiz-lara_2022,oria_2022}, which were grouped together using the distance threshold of \citetalias{lovdal_2022} but identified as separate structures using the more restrictive threshold of \citetalias{dodd_2023}. We found that with the threshold used by \citetalias{dodd_2023}, the purity of groups increases by a factor three compared to when the threshold of \citetalias{lovdal_2022} is used, while the group completeness does not change significantly. As such, our results go in the direction of the conclusion drawn by \citetalias{ruiz-lara_2022} that these structures are not related to each other. Moreover, since both Sequoia and L-RL64/Antaeus are retrograde, with average metallicities of [Fe/H]$=-1.47$ and [Fe/H]$=-1.67$, respectively \citep{bellazzini_2023,dodd_2023}, we can confirm that these structures are dominated by accreted particles. 

Regarding Sequoia, \citetalias{ruiz-lara_2022} found it to be composed by two clusters, the most significant of which has a significance of 9.29, justified by the clear overdensity it formed in the E-L$_z$ space. In our simulations, all clusters dominated by accreted particles with such a high significance are systematically formed by galaxies accreted less than 6~Gyr ago. Therefore, we suggest that the progenitor of Sequoia might have been accreted within the last 6~Gyr, which is more recent than the usually assumed timeframe \citep[$\simeq 9$~Gyr ago,][]{myeong_2019}. This contrasts with the accretion time of the progenitor galaxy at $z \sim 2$ that led to the Sequoia-like feature in the E-L$_z$ space found by \citet{garcia-bethencourt_2023} in their simulations. It has been suggested that Sequoia, along with other structures such as Arjuna, L'Itoi, and LMS-1, might be related to Gaia-Enceladus-Sausage based on their chemistry \citep{koppelman_2020,naidu_2020,naidu_2021}. However, it is largely accepted that the progenitor of Gaia-Enceladus-Sausage has been accreted 8-11~Gyr ago (\citealt{helmi_2018,dimatteo_2019,gallart_2019}, but see \citealt{donlon_2024}), at least 2~Gyr earlier than our estimation for the progenitor of Sequoia. Therefore, it is very unlikely that GES and Sequoia are the signatures of the same accreted galaxy, or that the Sequoia progenitor was a satellite of GES as proposed by \citet{naidu_2021}. Furthermore, our results clearly show that it can be hazardous to claim that structures are related together only based on the distribution of metallicity, in particular in the case of GES and Sequoia, since \citet{matsuno_2022} found that they both have different chemical trends using high-resolution spectroscopic measurement.

Regarding the Helmi stream \citep{helmi_1999}, it is a prograde structure with a median metallicity of [Fe/H]$=-1.52$ \citep{bellazzini_2023}. Following our results, we can unsurprisingly confirm its accreted nature. Furthermore, in \citetalias{lovdal_2022}, it is identified by two significant clusters, with the highest one having a significance of $S=9.83$. It is likely that this value is underestimated, as the Helmi stream is located in a region of the IoM space where the artificial haloes tend to be overpopulated (see Sect.~\ref{sec:backgen}). In all the cases, as for Sequoia, such a high significance suggests that its progenitor galaxy was accreted within the last $6~$Gyr. This accretion time is on the lower edge of the estimation obtained in previous works \citep[5-9~Gyr]{kepley_2007,koppelman_2019}. 

Another structure identified by both \citetalias{lovdal_2022} and \citetalias{dodd_2023} is Thamnos. It was first identified by \citet{koppelman_2019} as two distinct structures, Thamnos 1 and 2, in the IoM space. However, based on their dynamical and chemical properties, it has been suggested that these two structures likely originate from the same accreted galaxy, with a stellar mass of approximately $5 \times 10^6$~M$_\odot$ \citep{koppelman_2019,ruiz-lara_2022}. 
Both \citetalias{lovdal_2022} and \citetalias{dodd_2023} identify Thamnos as a group of several clusters linked under the same dynamical group. \citet{bellazzini_2023} measured a median metallicity of [Fe/H]$=-1.22$ for the stars belonging to the Thamnos group identified by \citetalias{lovdal_2022}, which is slightly higher than the values reported by \citetalias{ruiz-lara_2022} for Thamnos 1 and 2, of [Fe/H]$=-1.42$ and $-1.30$, respectively. This relatively high metallicity in the retrograde region suggests that Thamnos may not necessarily be dominated by accreted particles, but rather could have a substantial contribution from in-situ particles. Even if Thamnos is mainly composed of accreted stars, the contamination from in-situ particles could be significant, potentially around 40\%, as indicated by Fig.~\ref{fig:clLZage_feh_mg}. This high level of contamination by in-situ particles is in agreement with the recent estimation of \citet{dodd_2024}. Furthermore, in \citetalias{lovdal_2022}, the three clusters that they are located in a region of the IoM space where the significance can be overestimated due to the underpopulation of the artificial haloes in that region, although this is not the case in all the simulations. In the four simulations studied here, all the clusters found in that region of the IoM space have a similar significance between 3 and 3.5, and are populated between 30\% and up to 80\% of in-situ particles. Furthermore, even by putting aside the contribution of in-situ particles, most of these clusters are populated by particles from two progenitors, with a population ratio of $\sim 30\%$. The only exception is in Au.~23, where $75-80\%$ of the particles are from a single progenitor. Therefore, we caution against the assumption that Thamnos is a clear significant structure left by a single accreted galaxy. Instead, it may be a composite structure formed by several accreted galaxies and in-situ stars, which could explain its unique chemical properties that differ from any other substructures found in the stellar halo \citep{horta_2023}. At the current stage, we are not able to favour one possibility over the other, although it seems clear that the region of Thamnos is highly contaminated by in-situ particles. Further research, particularly focusing on the promising age-metallicity relation developed by \citet{dodd_2024}, and leveraging future data from Gaia, as well as upcoming large spectroscopic surveys such as DESI-MWS \citep{cooper_2023}, WEAVE \citep{jin_2024}, and 4-MOST \citep{dejong_2010}, will be essential to clarify the reality of the Thamnos structure.

Another point is regarding the L-RL3 structure (Cl.~3 of \citetalias{lovdal_2022}). It is a prograde cluster located at low energy with a median metallicity of [Fe/H]=-0.70 \citep{bellazzini_2023}, composed of 2,137 stars and a significance of 6.3. In all the simulations when in-situ particles are considered, the clusters with more than 1,000 particles are highly populated by in-situ particles (>40\%), and are usually dominated by them. Given this fact and considering its high median metallicity, we favour the hypothesis made by \citetalias{ruiz-lara_2022} that they are mostly populated by in-situ particles. Even by selection of the particles of the clusters with [Fe/H]$<-1.0$, we found that $\simeq 30\%$ of the particles are flagged as in-situ and that the accreted particles originate from two different progenitors with a contribution ratio to accreted particles of typically 60\% and 30\% respectively. As such, we suggest that L-RL3 is not a structure formed by a single accreted galaxy, and it is rather formed by the accumulation of in-situ and accreted particles. It is possible that L-RL3 is an artificial cluster created by the underpopulation of the artificial background haloes at its location.

Finally, our simulations tentatively suggest that well-defined structures in the E-$L_z$ space within the Solar vicinity, which are identifiable by eye, such as ED-2 \citep{dodd_2023,balbinot_2023}, ED-4/Typhoon \citep{tenachi_2022}, and possibly Shakti \citep{Malhan_2024}, may have been created by galaxies accreted relatively recently (within the last 6–7 Gyr). We find no evidence of similarly well-defined structures originating from older, even very massive, galaxies (see Fig.~\ref{fig:dominanceAccOnly}). However, these structures are assumed to be the debris left by accreted galaxies several order of magnitude less massive than the ones studied here. Therefore, it is possible that the current state-of-the art cosmological simulations such as Auriga are not able to resolve the dynamical imprint left by such small galaxies over several Gyr, although they are able to resolve dwarf galaxies and stellar streams down to 10$^6$ M$_\odot$ (see \citealt{grand_2024,riley_2024}, but also \citealt{panithanpaisal_2021,cunningham_2022,shipp_2022,horta_2024} for other cosmological suite of similar resolution). Furthermore, in these simulations, there remains uncertainty related to several aspects of the physics models, such as chemical enrichment, yields, and feedback processes, which could influence the detailed chemical patterns of the stellar debris. It will be interesting in the near future to study the survivability of these structures in the current simulations. 

Another potential caveat of our analysis concerns the possible presence of  misclassified in-situ stars that may actually have originated from accreted gas clouds. This misclassification could have several significant implications for the conclusions drawn from our study, depending on the actual ratio of misclassified in-situ stars. For instance, it might explain the high contamination by in-situ particles observed in Sect.~\ref{sec:recovrate}. Similarly, it could account for the predominance of in-situ particles in the majority of significant clusters, regardless of their orbital properties. This misclassification might also shed light on why clusters dominated by in-situ and accreted particles exhibit similar chemical properties and stellar formation ages, particularly in the case of retrograde clusters. However, in this context, if the fraction of misclassified in-situ particles is substantial, we would expect to observe differences in the distribution of stellar ages between misclassified in-situ particles and correctly identified accreted particles. Indeed, for a given accretion event, the accreted stars are expected to form before the gas is stripped from the accreted galaxy, while the misclassified in-situ stars would be born shortly afterward. Yet, in the identified significant clusters, we did not find that in-situ stars were systematically younger than the accreted ones. Identifying which in-situ particles might be misclassified is a challenging task within the {\sc Auriga} simulations due to the mesh-based hydrodynamical approach employed by {\sc AREPO} \citep{Weinberger_2020}, the code used to run these simulations. This approach does not directly allow for tracking the origin of gas. We plan to revisit the accreted/in-situ classification in the Auriga simulations in a future work dedicated to this task.

In conclusion, the accretion history of the Milky Way is still far from well established, with several structures identified in the stellar halo near the Solar vicinity potentially being misinterpreted as accreted or not originating from a single accretion event. Our results question the reality of several detected structures assumed to be formed by the disruption of an accreted galaxy. Multiple scenarios could explain the observed dynamical, chemical, and stellar age properties of these structures. The upcoming Gaia data release, along with medium and high-resolution spectroscopic observations from DESI-MWS \citep{cooper_2023}, WEAVE \citep{jin_2024}, and 4-MOST \citep{dejong_2010}, will significantly enhance our understanding of the accretion history of the MW by allowing us to probe more distant regions, less affected by in-situ stars and where the dynamical timescale, and so the surviving time of the dynamical debris left by accreted galaxies is longer \citep{binney_2008}. However, as this study demonstrates, it is crucial to compare these datasets and the methods used to analyse them with cosmologically motivated simulations. Such comparisons are essential for evaluating the effectiveness of the methods developed to identify these structures and for determining their origins within the broader context of galactic evolution.

\section*{Acknowledgements}

The authors want to thank C. Chiappini, T. Callingham, E. Dodd, A. Helmi, V. Hill, S. L\"ovdal, and D. Massari, for useful discussions, and comments, and their strong support. They also thank the anonymous referee for her/his careful reading of the paper and for her/his interesting comments and suggestions that helped improving its quality.
GFT and GB acknowledge support from the Agencia Estatal de Investigaci\'on del Ministerio de Ciencia en Innovaci\'on (AEI-MICIN) and the European Regional Development Fund (ERDF) under grant number PID2020-118778GB-I00/10.13039/501100011033 and the AEI under grant number CEX2019-000920-S. RG acknowledges financial support from an STFC Ernest Rutherford Fellowship (ST/W003643/1). This work uses simulations from the Auriga Project public data release \citep{grand_2024} available at \url{https://wwwmpa.mpa-garching.mpg.de/auriga/gaiamock.html}.

\onecolumn

\begin{landscape}
\begin{longtable}{c|c|c|c|c|c|c|c|c|c|c|c|c}
\caption{Parameters of the significant clusters identified in the kinematically selected halo in the Solar vicinity of Au.~27.}\\
\hline
\textbf{Id.} & \textbf{N$_\star$} & \textbf{Signi.} & \textbf{$\mathcal{F}_{acc}$} & \textbf{$[$Fe/H$]$} & \textbf{$[$Mg/Fe$]$} & \textbf{Age } & \textbf{En} & \textbf{L$_\mathrm{z}$ } & \textbf{L$_\perp$} & \textbf{Dom. Prog.} & $\mathcal{C}_{dp}$ & \textbf{Group} \\
 &  &  &  & &  & {$[$Gyr$]$} & {[$10^5$ km$^{2}$~s$^{-2}$]} & {[km~s$^{-1}$~kpc$^{-1}$]} & {[km~s$^{-1}$~kpc$^{-1}$]} & & \% &\\
\hline
\endfirsthead

\multicolumn{13}{c}{{\tablename} \thetable{} -- continued from previous page}\\
\hline
\textbf{Id.} & \textbf{N$_\star$} & \textbf{Signi.} & \textbf{$\mathcal{F}_{acc}$} & \textbf{$[$Fe/H$]$} & \textbf{$[$Mg/Fe$]$} & \textbf{Age } & \textbf{En} & \textbf{L$_\mathrm{z}$ } & \textbf{L$_\perp$} & \textbf{Dom. Prog.} & $\mathcal{C}_{dp}$ & \textbf{Group} \\
 &  &  &  & &  & {$[$Gyr$]$} & {[$10^5$ km$^{2}$~s$^{-2}$]} & {[km~s$^{-1}$~kpc$^{-1}$]} & {[km~s$^{-1}$~kpc$^{-1}$]} & & \% &\\
\hline
\endhead

\hline
\multicolumn{13}{r}{{Continued on next page}} \\
\endfoot

\hline
\endlastfoot
\label{tab:cl27}

1 & 484 & $3.3$ & 0.40 & $-0.79_{-0.34}^{+0.24}$ & $0.18_{-0.03}^{+0.06}$ & $10.1_{-0.03}^{+0.06}$ & $-1.73$ & $55$ & $714$ & In-Situ & $59.50$ &  \\
2 & 4228 & $5.3$ & 0.40 & $-0.76_{-0.33}^{+0.27}$ & $0.18_{-0.03}^{+0.06}$ & $10.0_{-0.03}^{+0.06}$ & $-1.75$ & $235$ & $241$ & In-Situ & $59.67$ & Orange \\
3 & 108 & $3.2$ & 0.26 & $-0.67_{-0.30}^{+0.29}$ & $0.16_{-0.02}^{+0.07}$ & $9.7_{-0.4}^{+1.7}$ & $-1.89$ & $333$ & $195$ & In-Situ & $74.07$ & Orange \\
4 & 19 & $3.3$ & 0.53 & $-1.03_{-0.27}^{+0.33}$ & $0.22_{-0.04}^{+0.04}$ & $10.8_{-0.9}^{+0.7}$ & $-1.70$ & $-1112$ & $131$ & In-Situ & $47.37$ &  \\
5 & 16 & $3.2$ & 0.44 & $-0.74_{-0.36}^{+0.07}$ & $0.13_{-0.06}^{+0.11}$ & $8.6_{-1.7}^{+3.4}$ & $-1.22$ & $1753$ & $478$ & In-Situ & $56.25$ & Cyan \\
6 & 481 & $3.7$ & 0.45 & $-0.80_{-0.38}^{+0.25}$ & $0.19_{-0.04}^{+0.05}$ & $10.2_{-0.8}^{+1.1}$ & $-1.64$ & $-104$ & $581$ & In-Situ & $54.68$ &  \\
7 & 68 & $4.5$ & 0.24 & $-0.70_{-0.18}^{+0.12}$ & $0.11_{-0.05}^{+0.09}$ & $8.2_{-2.5}^{+2.3}$ & $-1.33$ & $1561$ & $303$ & In-Situ & $76.47$ & Cyan \\
8 & 28 & $3.3$ & 0.11 & $-0.68_{-0.12}^{+0.11}$ & $0.08_{-0.02}^{+0.04}$ & $7.4_{-0.6}^{+1.4}$ & $-1.41$ & $1531$ & $191$ & In-Situ & $89.29$ & Cyan \\
9 & 1891 & $9.4$ & 0.27 & $-0.71_{-0.25}^{+0.19}$ & $0.13_{-0.06}^{+0.08}$ & $9.1_{-2.1}^{+1.9}$ & $-1.49$ & $1118$ & $292$ & In-Situ & $72.82$ & Cyan \\
10 & 117 & $3.0$ & 0.21 & $-0.70_{-0.21}^{+0.19}$ & $0.12_{-0.04}^{+0.07}$ & $8.7_{-1.8}^{+1.9}$ & $-1.50$ & $1180$ & $417$ & In-Situ & $78.63$ & Cyan \\
11 & 30 & $3.1$ & 0.33 & $-0.66_{-0.13}^{+0.20}$ & $0.17_{-0.02}^{+0.04}$ & $10.0_{-0.5}^{+0.9}$ & $-1.92$ & $260$ & $431$ & In-Situ & $66.67$ &  \\
12 & 38 & $3.4$ & 0.29 & $-0.73_{-0.24}^{+0.15}$ & $0.12_{-0.05}^{+0.11}$ & $8.7_{-2.3}^{+2.5}$ & $-1.44$ & $1392$ & $178$ & In-Situ & $71.05$ & Cyan \\
13 & 79 & $4.5$ & 0.16 & $-0.73_{-0.11}^{+0.10}$ & $0.10_{-0.04}^{+0.08}$ & $8.1_{-2.0}^{+2.0}$ & $-1.41$ & $1364$ & $613$ & In-Situ & $83.54$ & Cyan \\
14 & 240 & $7.5$ & 0.19 & $-0.72_{-0.14}^{+0.12}$ & $0.09_{-0.03}^{+0.08}$ & $7.9_{-1.9}^{+2.3}$ & $-1.34$ & $1536$ & $500$ & In-Situ & $80.83$ & Cyan \\
15 & 187 & $4.6$ & 0.27 & $-0.72_{-0.19}^{+0.10}$ & $0.11_{-0.04}^{+0.09}$ & $8.6_{-1.9}^{+2.0}$ & $-1.38$ & $1417$ & $759$ & In-Situ & $73.26$ & Cyan \\
16 & 21 & $3.0$ & 0.10 & $-0.71_{-0.08}^{+0.04}$ & $0.08_{-0.03}^{+0.06}$ & $7.3_{-1.5}^{+1.6}$ & $-1.42$ & $1376$ & $108$ & In-Situ & $90.48$ & Cyan \\
17 & 39 & $3.0$ & 0.38 & $-0.96_{-0.18}^{+0.27}$ & $0.19_{-0.03}^{+0.03}$ & $10.3_{-0.7}^{+1.0}$ & $-1.52$ & $-772$ & $1582$ & In-Situ & $61.54$ & Pink \\
18 & 42 & $3.2$ & 0.29 & $-0.76_{-0.23}^{+0.18}$ & $0.12_{-0.05}^{+0.06}$ & $8.9_{-2.0}^{+1.8}$ & $-1.26$ & $1619$ & $557$ & In-Situ & $71.43$ & Cyan \\
19 & 68 & $3.8$ & 0.19 & $-0.72_{-0.10}^{+0.12}$ & $0.09_{-0.03}^{+0.08}$ & $7.3_{-1.3}^{+3.2}$ & $-1.31$ & $1579$ & $712$ & In-Situ & $80.88$ & Cyan \\
20 & 238 & $7.5$ & 0.22 & $-0.72_{-0.15}^{+0.15}$ & $0.10_{-0.04}^{+0.07}$ & $7.9_{-1.5}^{+2.3}$ & $-1.28$ & $1669$ & $380$ & In-Situ & $78.15$ & Cyan \\
21 & 121 & $5.4$ & 0.25 & $-0.73_{-0.19}^{+0.15}$ & $0.11_{-0.05}^{+0.09}$ & $8.5_{-2.1}^{+2.4}$ & $-1.29$ & $1746$ & $729$ & In-Situ & $75.21$ & Cyan \\
22 & 169 & $3.5$ & 0.19 & $-0.73_{-0.12}^{+0.12}$ & $0.10_{-0.04}^{+0.08}$ & $7.7_{-1.9}^{+2.5}$ & $-1.32$ & $1568$ & $1108$ & In-Situ & $81.07$ & Cyan \\
23 & 55 & $3.9$ & 0.22 & $-0.71_{-0.15}^{+0.14}$ & $0.10_{-0.03}^{+0.08}$ & $7.6_{-1.7}^{+3.0}$ & $-1.21$ & $1917$ & $361$ & In-Situ & $78.18$ &  \\
24 & 96 & $3.4$ & 0.26 & $-0.91_{-0.36}^{+0.17}$ & $0.19_{-0.04}^{+0.07}$ & $10.2_{-0.8}^{+1.1}$ & $-1.51$ & $-389$ & $1670$ & In-Situ & $73.96$ & Pink \\
25 & 174 & $4.8$ & 0.27 & $-0.74_{-0.12}^{+0.18}$ & $0.11_{-0.05}^{+0.08}$ & $7.8_{-2.1}^{+2.8}$ & $-1.26$ & $1795$ & $1114$ & In-Situ & $72.99$ & Cyan \\
26 & 62 & $3.6$ & 0.16 & $-0.72_{-0.14}^{+0.15}$ & $0.10_{-0.05}^{+0.07}$ & $7.9_{-3.2}^{+2.1}$ & $-1.28$ & $1726$ & $910$ & In-Situ & $83.87$ & Cyan \\
27 & 112 & $3.1$ & 0.16 & $-0.72_{-0.09}^{+0.11}$ & $0.09_{-0.04}^{+0.08}$ & $7.0_{-1.4}^{+3.1}$ & $-1.27$ & $1733$ & $1334$ & In-Situ & $83.93$ & Cyan \\
28 & 60 & $3.1$ & 0.30 & $-0.76_{-0.15}^{+0.12}$ & $0.11_{-0.05}^{+0.09}$ & $8.8_{-1.9}^{+2.1}$ & $-1.28$ & $1228$ & $204$ & In-Situ & $70.00$ &  \\
29 & 190 & $3.7$ & 0.25 & $-0.89_{-0.33}^{+0.21}$ & $0.18_{-0.03}^{+0.06}$ & $10.0_{-0.6}^{+1.2}$ & $-1.48$ & $-133$ & $1863$ & In-Situ & $75.26$ & Pink \\
30 & 198 & $6.0$ & 0.23 & $-0.74_{-0.16}^{+0.12}$ & $0.10_{-0.05}^{+0.09}$ & $7.5_{-2.6}^{+3.0}$ & $-1.20$ & $1979$ & $667$ & In-Situ & $77.27$ & Cyan \\
31 & 133 & $6.1$ & 0.18 & $-0.70_{-0.14}^{+0.11}$ & $0.08_{-0.03}^{+0.08}$ & $7.0_{-1.8}^{+3.2}$ & $-1.27$ & $1797$ & $194$ & In-Situ & $81.95$ & Cyan \\
32 & 83 & $4.7$ & 0.23 & $-0.76_{-0.17}^{+0.10}$ & $0.10_{-0.04}^{+0.09}$ & $8.1_{-2.5}^{+2.6}$ & $-1.11$ & $2148$ & $459$ & In-Situ & $77.11$ & Cyan \\
33 & 55 & $3.8$ & 0.22 & $-0.70_{-0.14}^{+0.13}$ & $0.08_{-0.03}^{+0.09}$ & $6.9_{-2.3}^{+3.2}$ & $-1.16$ & $2126$ & $349$ & In-Situ & $78.18$ & Cyan \\
34 & 33 & $3.7$ & 0.30 & $-0.74_{-0.12}^{+0.13}$ & $0.13_{-0.07}^{+0.07}$ & $8.8_{-3.1}^{+2.3}$ & $-1.20$ & $1997$ & $144$ & In-Situ & $69.70$ & Cyan \\
35 & 58 & $4.2$ & 0.36 & $-0.72_{-0.15}^{+0.21}$ & $0.09_{-0.03}^{+0.09}$ & $7.3_{-3.1}^{+3.3}$ & $-1.11$ & $2329$ & $306$ & In-Situ & $63.79$ & Cyan \\
36 & 129 & $4.3$ & 0.30 & $-0.74_{-0.23}^{+0.14}$ & $0.10_{-0.05}^{+0.11}$ & $7.9_{-3.1}^{+3.3}$ & $-1.10$ & $2264$ & $734$ & In-Situ & $69.77$ & Cyan \\
37 & 27 & $3.1$ & 0.30 & $-0.79_{-0.27}^{+0.13}$ & $0.13_{-0.06}^{+0.10}$ & $9.1_{-2.5}^{+2.3}$ & $-0.95$ & $2707$ & $300$ & In-Situ & $70.37$ &  \\
38 & 62 & $3.1$ & 0.34 & $-0.93_{-0.28}^{+0.17}$ & $0.20_{-0.06}^{+0.04}$ & $10.0_{-0.8}^{+1.1}$ & $-1.10$ & $-286$ & $449$ & In-Situ & $66.13$ & Green  \\
39 & 89 & $4.2$ & 0.46 & $-1.06_{-0.45}^{+0.26}$ & $0.22_{-0.05}^{+0.04}$ & $10.5_{-1.1}^{+0.9}$ & $-1.09$ & $-879$ & $1563$ & In-Situ & $53.93$ & Blue \\
40 & 49 & $4.7$ & 0.51 & $-0.94_{-0.44}^{+0.18}$ & $0.20_{-0.04}^{+0.04}$ & $10.2_{-0.8}^{+1.4}$ & $-1.02$ & $-1164$ & $976$ & In-Situ & $48.98$ & Blue  \\
41 & 57 & $3.6$ & 0.53 & $-1.00_{-0.39}^{+0.24}$ & $0.20_{-0.04}^{+0.07}$ & $10.2_{-0.8}^{+1.7}$ & $-0.99$ & $-537$ & $263$ & In-Situ & $47.37$ & Green\\
42 & 36 & $4.2$ & 0.42 & $-0.99_{-0.28}^{+0.24}$ & $0.21_{-0.06}^{+0.03}$ & $10.0_{-0.7}^{+1.1}$ & $-1.00$ & $-650$ & $1807$ & In-Situ & $58.33$  & Blue \\
43 & 104 & $8.1$ & 0.52 & $-1.11_{-0.40}^{+0.29}$ & $0.21_{-0.03}^{+0.05}$ & $10.4_{-0.7}^{+0.8}$ & $-0.96$ & $-2100$ & $1733$ & In-Situ & $48.08$ & Blue  \\
44 & 75 & $3.3$ & 0.40 & $-1.05_{-0.33}^{+0.21}$ & $0.21_{-0.05}^{+0.05}$ & $10.2_{-0.6}^{+1.4}$ & $-1.10$ & $-1160$ & $2140$ & In-Situ & $60.00$ & Blue  \\
45 & 17 & $3.2$ & 0.47 & $-0.97_{-0.83}^{+0.12}$ & $0.18_{-0.03}^{+0.06}$ & $9.9_{-0.6}^{+1.6}$ & $-0.96$ & $-1345$ & $1809$ & In-Situ & $52.94$ & Blue  \\
46 & 55 & $6.2$ & 0.56 & $-0.93_{-0.42}^{+0.16}$ & $0.21_{-0.04}^{+0.03}$ & $10.2_{-0.8}^{+1.1}$ & $-0.97$ & $-1533$ & $1302$ & In-Situ & $43.64$ & Blue  \\
47 & 30 & $4.0$ & 0.43 & $-1.09_{-0.27}^{+0.18}$ & $0.20_{-0.04}^{+0.05}$ & $10.3_{-0.6}^{+0.9}$ & $-1.02$ & $-2076$ & $2059$ & In-Situ & $56.67$ & Blue  \\
48 & 18 & $3.4$ & 0.44 & $-1.05_{-0.24}^{+0.33}$ & $0.23_{-0.05}^{+0.04}$ & $10.5_{-0.7}^{+1.1}$ & $-0.94$ & $-1501$ & $1453$ & In-Situ & $55.56$ & Blue  \\
49 & 91 & $5.2$ & 0.54 & $-1.03_{-0.32}^{+0.22}$ & $0.20_{-0.04}^{+0.04}$ & $10.2_{-0.8}^{+0.9}$ & $-0.92$ & $-478$ & $979$ & In-Situ & $46.15$ & Blue  \\
50 & 41 & $3.6$ & 0.49 & $-1.05_{-0.23}^{+0.21}$ & $0.21_{-0.05}^{+0.04}$ & $10.3_{-1.0}^{+0.7}$ & $-0.95$ & $-488$ & $2518$ & In-Situ & $51.22$ & Blue  \\
51 & 36 & $4.6$ & 0.36 & $-1.01_{-0.40}^{+0.22}$ & $0.20_{-0.05}^{+0.06}$ & $10.2_{-0.7}^{+1.3}$ & $-0.99$ & $-1357$ & $2399$ & In-Situ & $63.89$ & Blue  \\
52 & 69 & $5.6$ & 0.43 & $-1.05_{-0.28}^{+0.26}$ & $0.21_{-0.05}^{+0.04}$ & $10.1_{-0.6}^{+1.0}$ & $-0.94$ & $-1066$ & $2102$ & In-Situ & $56.52$ & Blue  \\
53 & 12 & $3.3$ & 0.33 & $-1.02_{-0.62}^{+0.14}$ & $0.19_{-0.05}^{+0.04}$ & $9.9_{-1.3}^{+1.2}$ & $-0.94$ & $-2147$ & $2308$ & In-Situ & $66.67$ & Blue  \\
54 & 57 & $4.4$ & 0.47 & $-1.04_{-0.36}^{+0.33}$ & $0.20_{-0.04}^{+0.06}$ & $10.1_{-0.7}^{+1.3}$ & $-0.97$ & $-449$ & $2899$ & In-Situ & $52.63$ & Blue  \\
55 & 39 & $5.2$ & 0.44 & $-1.06_{-0.36}^{+0.25}$ & $0.20_{-0.03}^{+0.05}$ & $10.2_{-0.8}^{+0.9}$ & $-0.99$ & $-1068$ & $2758$ & In-Situ & $56.41$ & Blue  \\
56 & 26 & $3.1$ & 0.38 & $-1.00_{-0.21}^{+0.05}$ & $0.21_{-0.06}^{+0.05}$ & $10.2_{-0.9}^{+0.6}$ & $-0.93$ & $-177$ & $3152$ & In-Situ & $61.54$ & Blue  \\
57 & 15 & $3.1$ & 0.60 & $-0.92_{-0.18}^{+0.23}$ & $0.21_{-0.03}^{+0.02}$ & $10.1_{-0.2}^{+0.8}$ & $-0.85$ & $-1835$ & $839$ & Prog.1 & $46.67$ \\
58 & 14 & $3.4$ & 0.79 & $-1.34_{-0.43}^{+0.39}$ & $0.22_{-0.05}^{+0.03}$ & $11.0_{-1.2}^{+0.7}$ & $-0.87$ & $-1642$ & $2318$ & Prog.1 & $42.86$ & Blue  \\
59 & 26 & $3.4$ & 0.73 & $-1.28_{-0.45}^{+0.37}$ & $0.21_{-0.02}^{+0.04}$ & $10.6_{-0.6}^{+1.7}$ & $-0.80$ & $-695$ & $1527$ & Prog.1 & $50.00$ \\
60 & 15 & $3.4$ & 0.53 & $-1.05_{-0.33}^{+0.21}$ & $0.20_{-0.06}^{+0.02}$ & $10.1_{-2.8}^{+0.9}$ & $-0.79$ & $-1531$ & $1683$ & In-Situ & $46.67$ & Blue  \\
61 & 385 & $17.2$ & 0.85 & $-0.85_{-0.46}^{+0.43}$ & $0.18_{-0.06}^{+0.05}$ & $7.8_{-3.7}^{+2.9}$ & $-0.74$ & $-2834$ & $1948$ & Prog.4 & $72.99$ & Blue  \\
62 & 12 & $3.0$ & 0.67 & $-1.02_{-0.20}^{+0.14}$ & $0.20_{-0.04}^{+0.06}$ & $10.0_{-3.5}^{+0.5}$ & $-0.76$ & $-2297$ & $2450$ & Prog.1 & $33.33$ & Blue  \\
63 & 11 & $3.2$ & 0.82 & $-1.18_{-0.17}^{+0.19}$ & $0.21_{-0.02}^{+0.02}$ & $7.6_{-0.4}^{+1.3}$ & $-0.13$ & $841$ & $522$ & Prog.6 & $81.82$ &  \\
\hline
\end{longtable}
\tablefoot{For each cluster, Col.~1 presents its Id., Col.~2 the number of stellar particles belonging to it, Col.~3 its significance, Col.~4 the fraction of accreted particles composing it, Col.~5 its median metallicity, Col.~6 its median Mg abundance, Col.~7 the median age of its stellar particles, Col.~8 its mean total energy, Col.~9 its mean vertical angular momentum, Col.~10 its mean perpendicular angular momentum, Col.~11 the dominant progenitor, Col.~12 the contribution (in percentage) of the dominant progenitor to the cluster population, and Col.~13 the group of clusters to which it belongs, identified by the same colour as used in the dendrogram of Fig.\ref{fig:dendrogram} to identify the groups. The uncertainties quoted in Cols.~5-7 correspond to the 16th and 84th percentiles of the distribution.}
\end{landscape}

\bibliographystyle{aa}
\bibliography{./biblio}

\end{document}